\documentclass[apj]{emulateapj}

\usepackage{graphicx}
\usepackage{txfonts}
\usepackage{lscape}
\usepackage{aas_macros}
\usepackage{amssymb}
\usepackage{natbib}
\bibpunct{(}{)}{;}{a}{,}{,}
\usepackage{longtable}
\usepackage[hang,bf,footnotesize]{subfigure}
\usepackage[figuresright]{rotating}

\begin{document}

\title{Chandra/ACIS-I study of the X-ray properties of the NGC~6611 and M16 stellar population}

\author{M. G. Guarcello\altaffilmark{1}, M. Caramazza\altaffilmark{2}, G. Micela\altaffilmark{2}, S. Sciortino\altaffilmark{2}, J. J. Drake\altaffilmark{1}, L. Prisinzano\altaffilmark{2}}

\altaffiltext{1}{Smithsonian Astrophysical Observatory, MS-67, 60 Garden Street, Cambridge, MA 02138, USA}
\altaffiltext{2}{INAF - Osservatorio Astronomico di Palermo, Piazza del Parlamento 1, 90134 Palermo, Italy}

\begin{abstract}
Mechanisms regulating the origin of X-rays in YSOs and the correlation with their evolutionary stage are under debate. Studies of the X-ray properties in young clusters allow to understand these mechanisms. One ideal target for this analysis is the Eagle Nebula (M16), with its central cluster NGC6611. At $1750\,pc$ from the Sun, it harbors 93 OB stars, together with a population of low-mass stars from embedded protostars to disk-less Class~III objects, with age $\leq3\,Myrs$. We study an archival $78\,ksec$ Chandra/ACIS-I observation of NGC6611, and two new $80\,ksec$ observations of the outer region of M16, one centered on the Column~V, and one on a region of the molecular cloud with ongoing star-formation. We detect 1755 point sources, with 1183 candidate cluster members (219 disk-bearing and 964 disk-less). We study the global X-ray properties of M16 and compare them with those of the Orion Nebula Cluster. We also compare the level of X-ray emission of Class~II and Class~III stars, and analyze the X-ray spectral properties of OB stars. Our study supports the lower level of X-ray activity for the disk-bearing stars with respect to the disk-less members. The X-ray Luminosity Function (XLF) of M16 is similar to that of Orion, supporting the universality of the XLF in young clusters. 85\% of the O stars of NGC6611 have been detected in X-rays. With only one possible exception, they show soft spectra with no hard component, indicating that mechanisms for the production of hard X-ray emission in O stars are not operating in NGC~6611.
\end{abstract}



\section{INTRODUCTION}

	Since the discovery of the intense X-ray emission from young pre-Main Sequence (PMS) stars \citep{Fei81,Mon83}, X-ray observations of active star-forming regions and young galactic clusters become an efficient method to study the star formation process and the properties of young stars. In fact, the level of X-ray emission in PMS stars, higher than that of field main-sequence stars, provides a very efficient means of selecting stars associated with star forming regions and young clusters. In the last decade a large number of young clusters have been observed with X-ray telescopes, such as the {\it Chandra X-ray Telescope} \citep{Weis02} and {\it XMM-Newton} \citep{Jans01}, in order to select their young members and study their X-ray activity. To the present day, the longest X-ray observations of star-forming regions are the $839\,ksec$ Chandra Orion Ultradeep Project (COUP, \citealp{Get05}), the $1.08\,Msec$ Chandra Cygnus~OB2 Legacy Survey \citep{Dra09} and the $1.6\,Msec$ Chandra Carina Complex Project (CCCP, \citealp{Tow11}). \par
While we have a reasonably detailed comprehension of the coronal activity and accretion phenomenon in low-mass pre-main sequence stars, and of the mechanisms for X-ray emission in massive stars, several key topics are still not completely understood. For instance, how the presence of a circumstellar disk affects X-ray activity \citep{Fla03}, the nature of the huge spread in X-ray activity observed in almost all the young clusters \citep{Fei02}, and the importance of the proposed mechanisms for hard X-ray emission in massive stars \citep{Bab97}. \par
	In this paper we address some of these topics by studying the global X-ray properties of the young stars associated with the Eagle Nebula (M16) and its central cluster NGC~6611. Situated in the Sagittarius arm and lying in the southern sky, this cluster is $1750\,pc$ away from the Sun. It suffers only a low average extinction in the central cavity of the molecular cloud cleared by the cluster itself ($A_V\sim2.6^m$, \citealp{io07}), but becomes more absorbed in the surrounding area where the cloud is still dense \citep{io10}. The central cluster hosts 93 OB stars \citep{Hil93}, including one of the rare candidate O4 stars observed in our galaxy (W205, with a mass of $75-80\,M_{\odot}$, \citealt{Eva05}), the massive binary system W175, composed of an O5V star \citep{Eva05} and a late-O (likely O8.5) star \citep{Mar08}, some Herbig Ae/Be stars and the two magnetic stars W080 (B1V) and W601 (B1.5); \citet{Mar08,Alec08}. \par 
The region is also rich in Pre-Main Sequence (PMS) stars with low- and intermediate-mass. In our previous studies of this region \citep{io07,io09,io10b,io10}, we detected a total of 1937 candidate PMS stars associated with M16. The disk-less stars have been selected from their intense X-ray emission, and the disk-bearing stars from their excesses in infrared bands. The X-ray selected young stars of NGC~6611 have a median age of $\sim1\,Myr$ \citep{io07}, but in the whole cloud a star formation sequence has been proposed \citep{io10}, with the oldest star formation events to the south-east ($\sim 3\,Myr$) and the youngest in the north-west ($<1\,Myr$). Together with these large populations of Class~II and Class~III objects, a significant number of embedded Class~0/I sources \citep{Ind07}, water masers \citep{Hea04}, and candidate Herbig-Haro objects \citep{Mea86}, have been identified in sites of M16 with ongoing star-formation activity, such as the structures called ``the Pillars of Creation'' and ``Column~V'' \citep{Hes96,Mcc02,Sug07}, the Bright Rimmed Cloud (BRC) SFO30 northward of the cluster, and a young cluster in the north-east largely embedded in the molecular cloud \citep{Ind07,io09}. \par 
	The Eagle Nebula hosts a large variety of young X-ray emitting sources whose X-ray properties will be studied in detail in this paper. Our scope is to characterize the global X-ray properties of the cluster, to study how X-ray activity of cluster members is affected by the presence of the disks, and finally to take advantage of the large massive-star population of NGC~6611 to study X-ray emission in OB stars. The paper is organized as follows: in Sect. \ref{analysis_sect} we review the data analysis, source detection, the identification of the stellar counterparts of the detected X-ray sources, and the spectral analysis. The distributions of plasma temperature and hydrogen column density of the M16 stars are analyzed in Sect. \ref{xpro_sec}. The X-ray luminosity function of the whole population, as a function of the stellar mass and of evolutionary status, is studied in Sect. \ref{lx_sec}; in Sect. \ref{mass_sec} we study the X-ray emission of massive stars and in Sect. \ref{regions_sec} we review the X-ray properties of the stellar population of selected regions of the Eagle Nebula. In Appendix \ref{cata_sec} we present the X-ray catalog of detected sources.
 

\section{Data analysis and catalog}
\label{analysis_sect}
	\subsection{Sources identification and photon extraction}
	\label{phot_extr_sect}

       \begin{figure*}[]
        \centering
        \includegraphics[width=16cm]{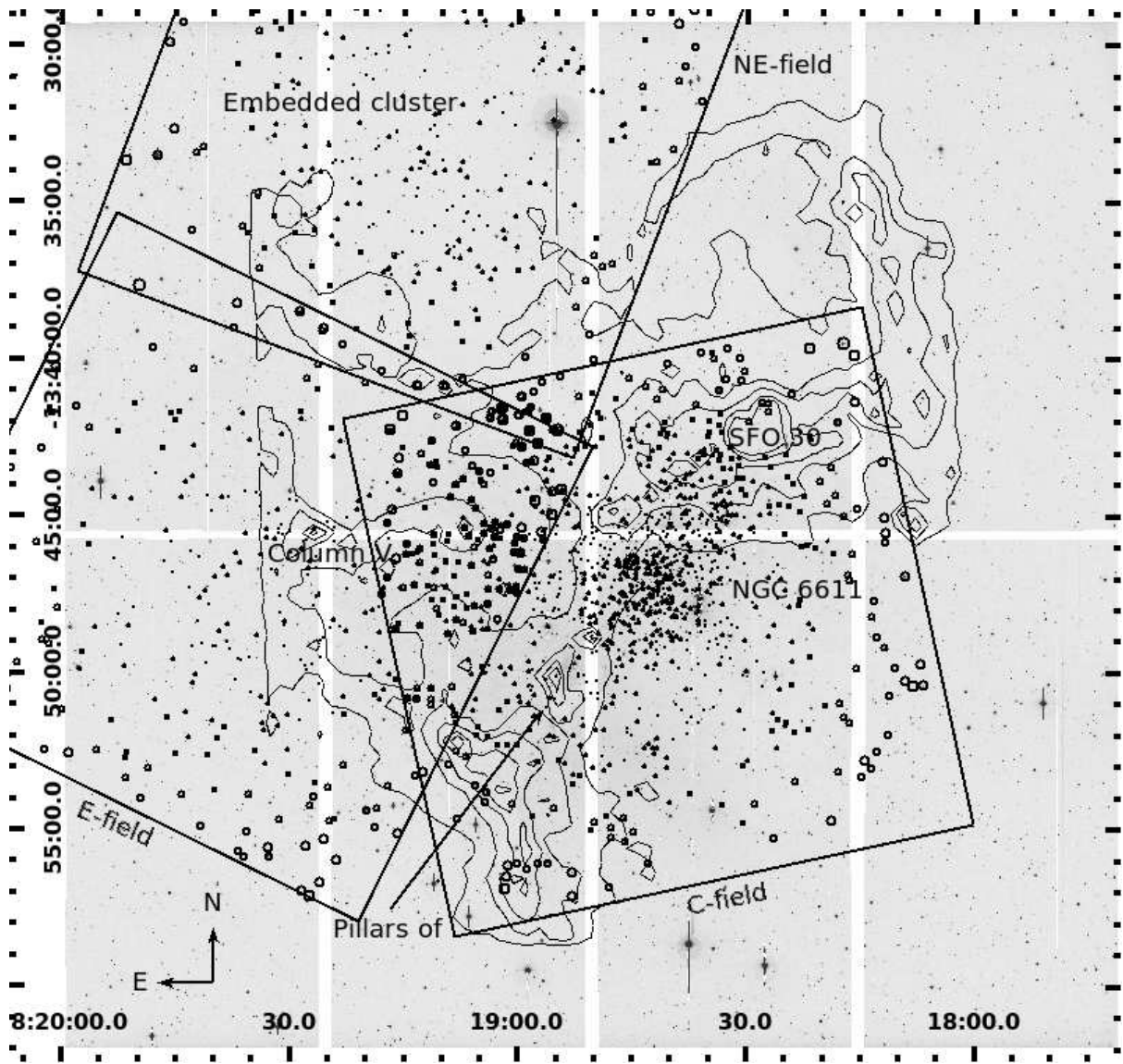}
        \caption{Image in I band of M16, with the contours encompassing the emission at $8.0\,\mu m$ obtained from Spitzer/IRAC. The rotated squares are the Chandra/ACIS-I fields of the observations analyzed in this work. The black circles mark the positions of detected X-ray sources. The radius of each circle is proportional to the PSF area at sources position. The positions of Pillars of Creation, Column~V, the SFO30 cloud, the North-East embedded cluster, and NGC~6611 are also shown.}
        \label{fov_image}
        \end{figure*}

Fig. \ref{fov_image} shows the image of the Eagle Nebula in the I band obtained with the Wield Field Imager (WFI) mounted on the 2.2m telescope at ESO \citep{io07}. The contours mark the dust emission detected at $8.0\,\mu m$ with Spitzer/IRAC \citep{Ind07,io09}, marking the Pillars of Creation, the Column~V and the SFO30 cloud (the outer contours delimit the size of the used IRAC image). The rotated squares represent the fields observed with Chandra/ACIS-I. The archival observation of $78\,ksec$ \citep{Lin07} is centered on NGC~6611 (the $c-field$) and it includes the Pillars of Creation and the SFO30 cloud; the two new observations of $80\,ksec$ (P.I. Guarcello) are centered on Column~V (the $e-field$), and on a group of Class~I and embedded Class~II stars (the $ne-field$) which is younger than $1\,Myr$. \par
	X-ray image reduction and source detection in all the fields are described in \citet{io10}, where detected X-ray sources (1158, 363 and 315 in the $c-$, $e-$, and $ne-fields$, respectively) with their optical-NIR counterparts have been used to classify YSOs in this region. Briefly, for each observation a {\it Level 2} event file has been obtained using the CIAO 4.0\footnote{http://cxc/harvard.edu/ciao} tool {\it acis-process-events}, retaining only the events incompatible with cosmic rays and removing the $0.5^{\prime \prime}$ event position randomization added by the standard Chandra data processing pipeline. Source detection has been performed with PWDetect\footnote{http://www.astropa.unipa.it/progetti\_ricerca/PWDetect} \citep{Dami97} adopting a threshold corresponding to 10 spurious detections. The total number of detected X-ray sources is 1755 (81 sources are in the overlapping region between the $c-field$ and the $e-field$). As described in \citet{io10}, the mass limit of the X-ray sources with optical counterparts (for which it was possible to estimate the mass) is equal to $0.2\,M_{\odot}$. Fig. \ref{fov_image} shows the photon extraction regions of each detected source. The clustering of sources in NGC~6611 is evident, but this image clearly shows that also the outer parts of the nebula are well populated by young stars. \par 
Photon extraction has been made using the IDL software ACIS Extract\footnote{http://www2.astro.psu.edu/xray/docs/TARA/ae\_users\_guide.html} ($AE$, \citealt{Bro10}), which uses TARA\footnote{http://www.astro.psu.edu/xray/docs/TARA}, CIAO, FTOOLS\footnote{http://heasarc.gsfc.nasa.gov/docs/software/ftools} and MARX\footnote{http://space.mit.edu/CXC/MARX/} packages. AE calculates the Point Spread Function (PSF) for each source, using it to define a photon extraction region as the region encompassing 90\% of the PSF evaluated at $1.49\,keV$. Background events have been extracted for each source in circular annuli centered on the sources, applying a mask for source counts defined in two iterations (first a circular area covering 99\% of the local PSF and then with a more accurate mask region). The extraction regions of crowded sources are reduced in order to not overlap each other, down to 40\% of the local PSF in the most crowded regions. In this first phase, AE uses the source positions provided by the source detection algorithm (PWDetect in our case). Once both the PSF and the background for each source are evaluated, AE computes new source positions by correlating the source images with the local PSF model. The user has the possibility to use these new positions or retain the set of positions provided by the source detections algorithm. As suggested by the AE manual, we updated the positions of the sources observed with an off-axis larger than $5^{\prime}$ (a further coordinates correction of $\Delta_{\alpha}=-0.18^{\prime \prime}$ and $\Delta_{\sigma}=-0.05^{\prime \prime}$ has been applied to cross-correlate the X-ray catalog with the optical-infrared catalog, see Sect. \ref{multi_cat_sec}). AE, then, repeats the computation of the PSF and the background and the photon extraction using the new positions. For each source, AE provides information such as the net observed counts, the light curves and a probability that the observed light curve is constant based on a Kormogorov-Smirnov test; the most relevant are summarized in the electronic catalog described in Appendix \ref{cata_sec}. \par

	\subsection{Stellar counterparts of the X-ray sources}
	\label{multi_cat_sec}

Detected X-ray sources are classified in \citet{io10} according to the properties of their optical/infrared counterparts. Among the 834 sources with disks, selected by their NIR excesses with respect to the expected photospheric emission, 219 have an X-ray counterpart and are classified as {\it ``disk-bearing members''}. We classified as {\it ``disk-less member''} 964 X-ray sources that are younger than $10\,Myr$ and with an extinction higher than $2.6^m$, according to their position in the optical color-magnitude diagram. A total of 76 sources are older than $10\,Myr$ and/or have extinction smaller than $2.6^m$. These sources are most likely foreground main sequence stars detected in X-rays, and then they have been classified as {\it ``foreground sources''}. A total of 504 X-ray sources are not identified with a known stellar counterpart. In \citet{io10} we compared our count-rate limit ($6.25\times10^{-5}\,counts/s$), properly converted into a limit flux in the $0.5-10\,keV$ energy band, with the $Log\,N$ vs. $Log\,S$ distribution of extragalactic sources in the ELAIS field shown in \citet{Puccetti2006}, estimating an upper limit of the observed extragalactic X-ray sources equal to 193.5. We then expect that about 300 X-ray sources without optical-infrared counterparts can be very embedded sources associated with M16 (a significant fraction of them lie in the regions with high extinction, such as the $ne-field$) or background sources lying in our Galaxy. Since we cannot discern among these hypotheses, we simply classified them as {\it ``background sources''} and we have not considered them in our study of the global X-ray properties of the M16 population.\par
\begin{figure*}[]
        \centering
        \includegraphics[width=6.5cm]{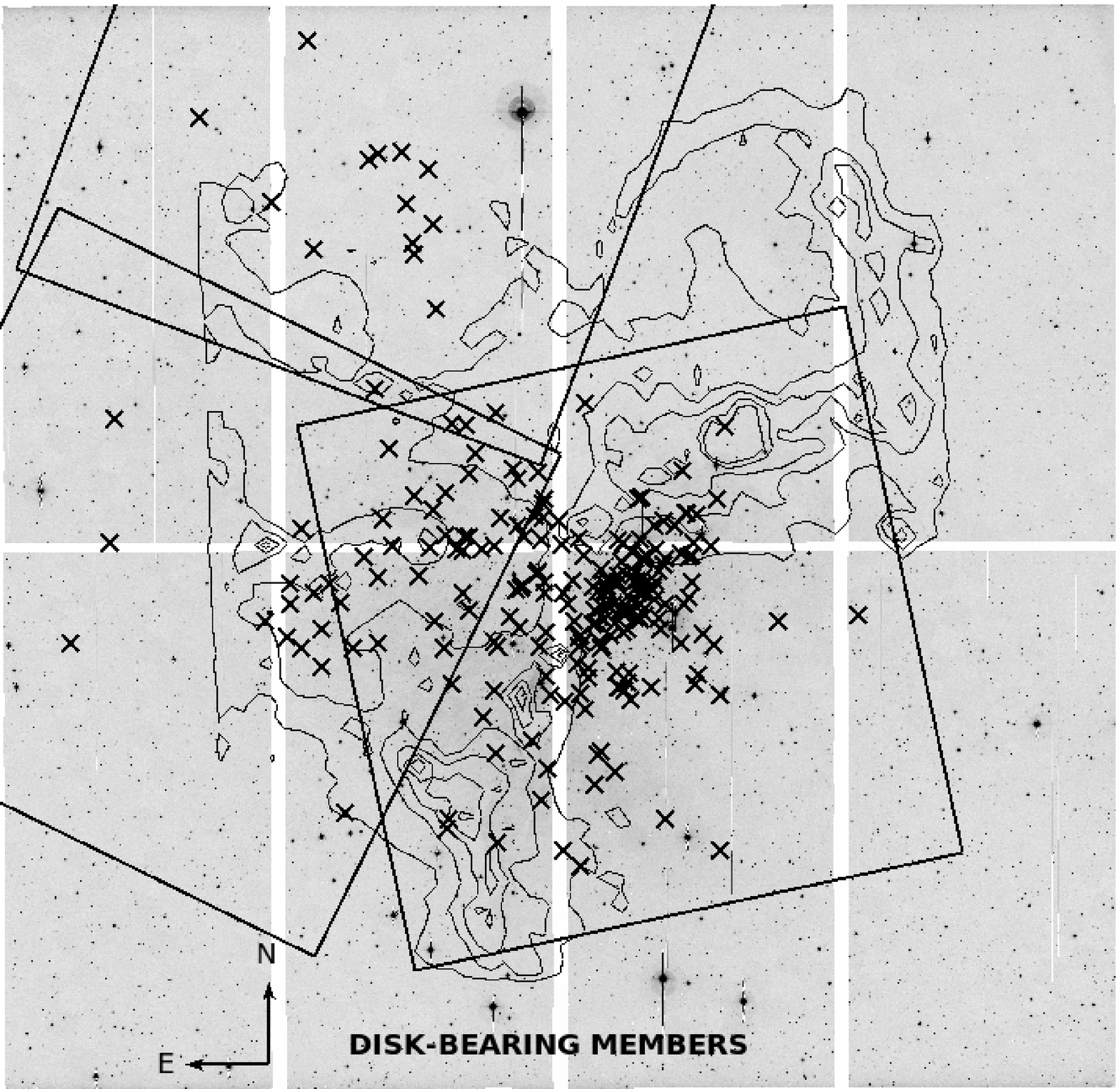}
        \includegraphics[width=6.5cm]{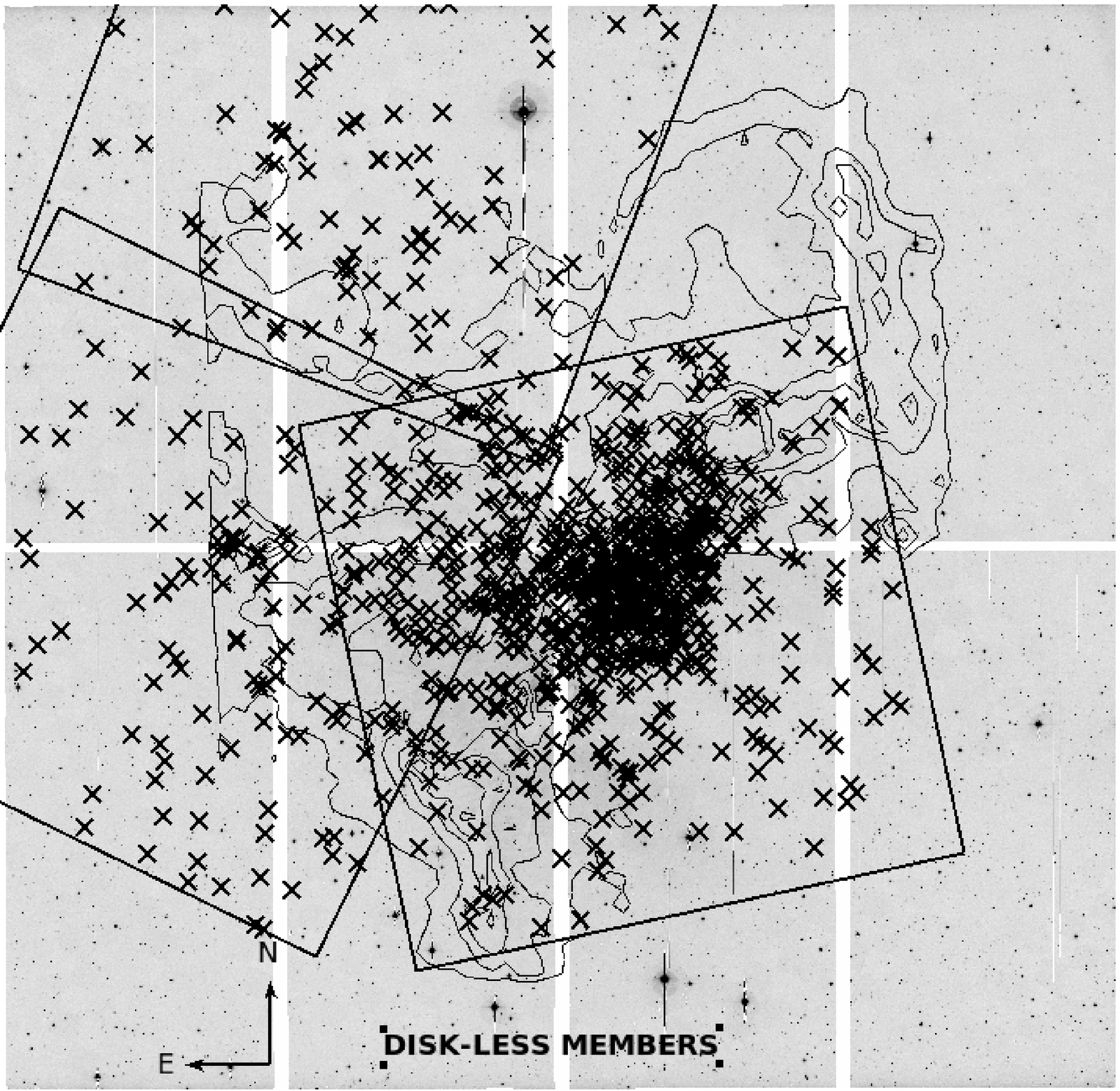}
        \includegraphics[width=6.5cm]{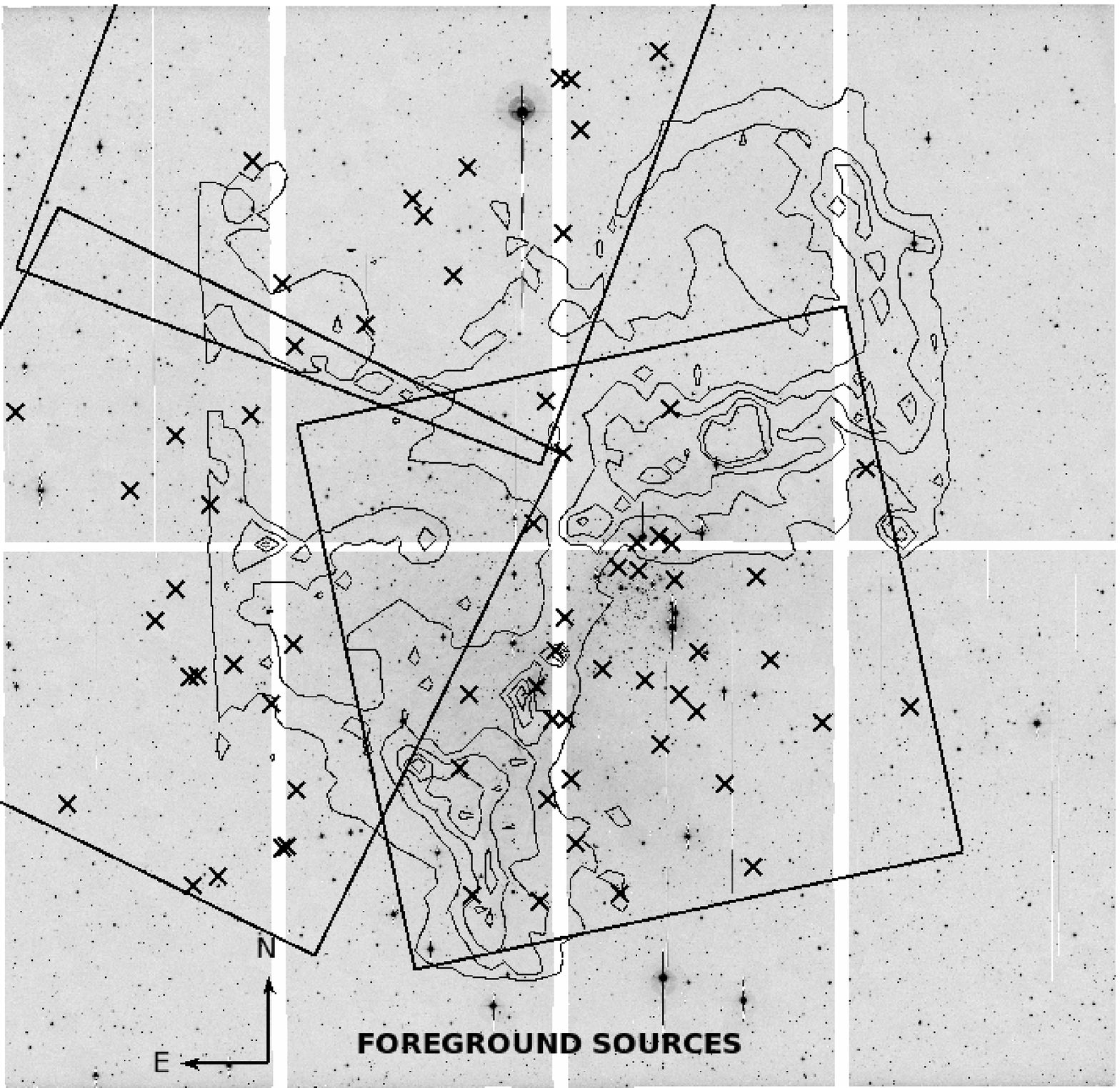}
        \includegraphics[width=6.5cm]{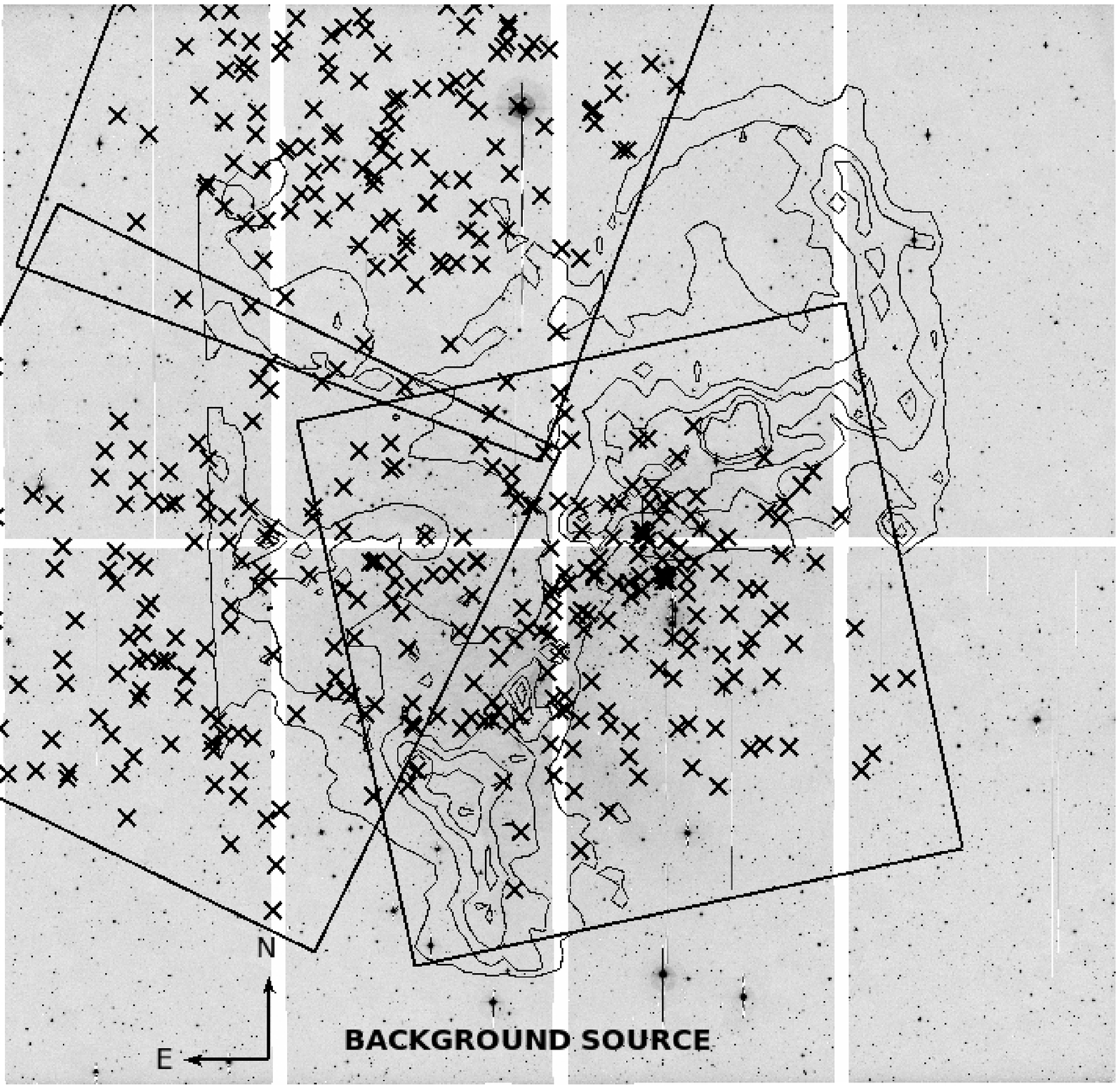}
        \caption{Spatial distribution of the four classes of X-ray sources discusses in Sect. \ref{multi_cat_sec} (disk-bearing and disk-less members, foreground and background sources), with overplotted the contours from the [8.0]  and the ACIS field used in Fig. \ref{fov_image}.}
        \label{classispadis_image}
        \end{figure*}

	Fig. \ref{classispadis_image} shows the spatial distributions of these four classes of X-ray sources. The disk-bearing stars are clearly clumped in NGC~6611, toward the Column~V and in the embedded north-east cluster. The disk-less members are more numerous and  distributed across the whole cloud. Also the background sources are sparse in the field, but the high concentration of them in the regions with the highest extinction and embedded ongoing star formation activity (the $ne-field$, eastward the Column~V, and in the tip of the Pillars of Creations and Column~V) is consistent with the idea that most of these sources are very embedded young stars associated with M16. \par

	\subsection{Spectral analysis: spectral fitting}
	\label{spectra_sec}

	The determination of the absorption corrected X-ray luminosity ($L_X$), as well as the plasma temperature ($kT$) and hydrogen column density ($N_H$), requires the analysis of the X-ray spectra. AE provides both the source and background spectra, the {\it redistribution matrix files} (RMF) and the {\it ancillary response files} (ARF). We fit the observed spectra with thermal plasma (with both one and two temperatures) and power-law models. We use the APEC ionization-equilibrium thermal plasma code \citep{Smi01}, assuming the sub-solar elemental abundances of \citet{Mag07}; the absorption was treated using the WABS model \citep{Mor83}. The one-temperature thermal model was applied to all the sources with more than 25 counts; while the two temperature thermal model was applied to each source with more than 80 counts. The power-law model has been applied to those sources with hard spectra for which the best-fit thermal model predicts a plasma temperature $kT\,>\,5\,keV$. In order to avoid false convergences due to local minima in the $\chi^2$ space, we applied different sets of initial values to each model. When more than one model has been used for a given source, we choose the best model by the $\chi^2$ probability and visual inspection of the spectrum. \par
In the $c-field$ we were able to obtain a good spectral fit for 441 sources, 114 in the $e-field$ and 60 in the $ne-field$ (out of a total of 1158, 363 and 315 sources, respectively), for a total of 575 sources with a good spectral fit (40 sources in the overlapping regions). For 33 sources the best fit model is a two temperature thermal plasma model. For a total of 20 sources the best fit model is a power-law spectrum: 15 of them are classified as {\it ``background sources''}, the remaining 5 as {\it ``disk-less members''}. \par  

	\subsection{Spectral analysis: energy quantiles}
	\label{quant_sec}

	Spectral analysis cannot be performed for faint sources. For them, an estimate of the X-ray luminosity can be provided by the analysis of the photon energy quantiles \citep{Hon04}. This method is similar to the analysis of the X-ray hardness ratio, adopted in several works, but it does not suffer the drawback of the choice of the energy bands used to define the ``X-ray colors''. Adopting the definition by \citet{Hon04}, if $x\%$ is the percentage of the total counts with energy below the value $E_{x\%}$, the quantile $Q_x$ is defined as: \par

\begin{equation}
Q_x=\frac{E_{x\%}-E_{lo}}{E_{up}-E_{lo}}
\end{equation}

	where $E_{lo}$ and $E_{up}$ are the lower and upper limits of the energy range, respectively $0.5\,keV$ and $8.0\,keV$ in our case. We used the median energy (i.e. the 50\% quartile) and the 25\% and 75\% quartiles, which are combined in the following independent variables: $log\left(Q_{50}/\left( 1-Q_{50} \right) \right)$ and $3\times \left(Q_{25}/Q_{75} \right)$. In order to estimate the values of $N_H$ and $kT$ for each source, the observed quantile variables have been compared (by interpolation) with a theoretical grid of thermal models with sub-solar abundances and a grid of power-law models with spectral indices typical of AGN \citep{Bra01}. \par

        \begin{figure*}[]
        \centering
        \includegraphics[width=14cm]{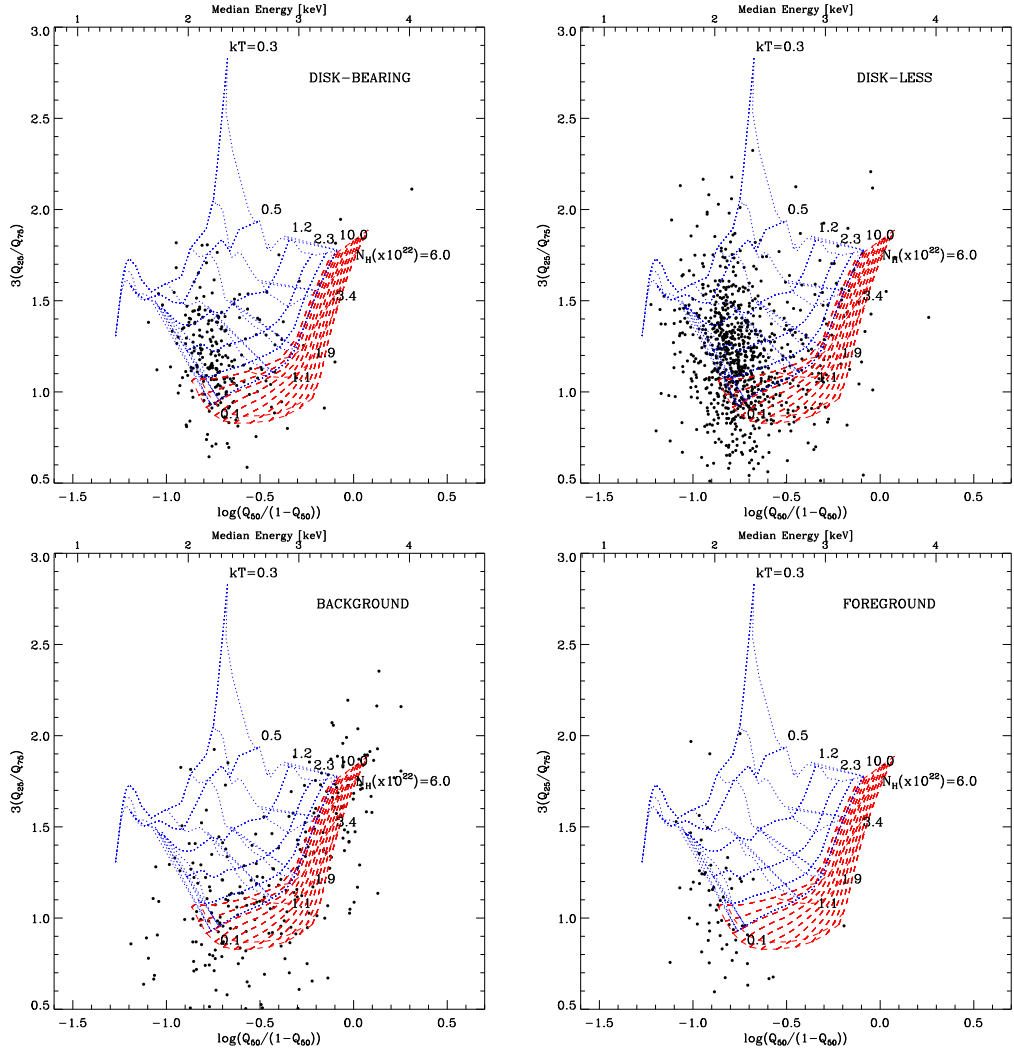}
        \caption{Quantiles of the four classes of X-ray sources, with overplotted the grids derived from thermal one-temperature spectra (blue lines) and power-law spectra (red lines). Some of the values of $N_H$ and $kT$ for the thermal models are shown. The power law models have indices ranging from 0 to 2.0 and the same $N_H$ of the thermal models.}
        \label{q_image}
        \end{figure*}

		Fig. \ref{q_image} shows the thermal and power-law grids and the quantiles of the X-ray sources (only the sources with a significance larger than 2 are shown; see Appendix \ref{cata_sec}.) for the four groups of sources defined in Sect. \ref{multi_cat_sec}. The thermal models have a plasma temperature ranging from $0.3\,keV$ to $10\,keV$ and hydrogen column density from $0.01\times10^{22}\,cm^{-2}$ to $6\times10^{22}\,cm^{-2}$; the power laws have the same hydrogen column density range and a photon index ranging from 0.1 to 2.0. The actual grids used for the interpolations are finer than those shown in Fig. \ref{q_image}. \par
The disk-bearing and disk-less sources (upper panels) are mostly concentrated in the region with $1.2\,keV \leq kT \leq 4\,keV$ and from $0.1\times10^{22}\,cm^{-2} \leq N_H \leq 1.1\times10^{22}\,cm^{-2}$, corresponding to a wide range of visual extinction. Both the background and the foreground sources are mostly outside the thermal grid. The distribution of the latter suggest values of $N_H$ smaller than those used for the grid. The former mostly show harder spectra and, in most cases compatible with power-law models. However, from the grid it is not possible to discriminate the real extragalactic sources from the embedded cluster members, given the degeneracy between very hard thermal spectra and power laws. \par
	A close inspection of Fig. \ref{q_image} shows the presence of 8 background sources in the area of the soft thermal spectra corresponding with $kT<0.5\,keV$. Even if they are not very bright (the median net counts of this sample is 19), the probability associated with the hypothesis that they are spurious detections is negligible (see the $PBS$ parameter, explained in Appendix \ref{cata_sec}). Five of these sources lie in the $c-field$ and have a median photon energy between $1.3\,keV$ and $1.8\,keV$. It is not obvious what these sources are. One intriguing possibility is that some of them are compact objects associated with M16, which are remnants of the explosion of a massive star. The possibility that M16 hosted a supernova explosion in the past has been suggested by \citet{Flag11} to explain the hot dust ($\sim70\,K$) they observed in the cloud. However, NGC~6611 is not old enough to firmly claim that the most massive members already exploded as supernovae, and the median photons energy of these 8 sources is higher than that of the candidate neutron stars recently identified in the Carina complex by \citet{Town11}. \par
In total, 846 sources lie inside the thermal grid. For these sources the individual values of $N_H$ and $kT$ have been calculated by interpolating their quantile variables in the thermal grid, and then their X-ray luminosity, $L_X$, using the distance adopted for the cluster ($1750\,pc$, \citealt{io07}) and correcting for the individual absorption. To estimate the uncertainties in the X-ray stellar parameters from the grid, we calculated the errors in their photons energy quantiles, following the procedure of \citet{Hon04} and propagated to the independent quantile quantities. Then, we repeated the interpolation with the grid four times, each time with a different combination of sum and subtraction of the errors to the quantile quantities. In this way, we had, for each star and for each parameter, 4 slightly different values. For each nominal value (i.e. those derived from the grid without adding or subtracting the errors) of the three quantities, the upper error has been defined as the difference between the maximum of this 4 measures and the nominal value, the lower error as the difference with the minimum value. The median error for the three parameters are $\sigma_{N_H}=0.7\,cm^{-2}$, $\sigma_{k_T}=0.4\,keV$, and $\sigma_{log \left(L_X \right)}=0.3\,erg/s$. \par
	A total of 312 sources lie inside to the power-law grid, and, as shown in Fig. \ref{q_image}, 82\% of them are background sources (with a wide range of hydrogen column density) and disk-less stars. There are 346 sources lying outside both grids. Most of them have $3Q_{25}/Q_{75}$ smaller than those of the grid models. \citet{Hon04} suggested that low significance sources can fall in the area of the diagram with $3Q_{25}/Q_{75}<0.8$. In fact the sources in this region of the diagram have a median of 12 counts, indicating they mostly fall in this region because of low-counts. \par 

	\subsection{Upper limits}
	\label{uplim_sec}
 
	The main aim of this paper includes the comparison between the X-ray properties of disk-less and disk-bearing stars, the study of the global X-ray properties of the M16 population, and their comparison with those of the Orion members. A reliable comparison requires the knowledge of the upper limit of X-ray luminosity for the undetected members. The problem of the incompleteness of the sample of Class~III objects of M16 will be discussed later, since we do not know the X-ray undetected Class~III stars. This is not the case for the disk-bearing members, since we have detected in X-rays 219 sources with infrared excesses out of a total of 834 previously identified. The upper limit of the photon count rate at the positions of the undetected disk-bearing stars is calculated with PWDetect, adopting the same threshold significance used for source detection. We have converted these upper limit count rates to X-ray luminosity using as conversion factor the median of the ratio between the X-ray luminosities and photon count rates for the detected members. \par

\section{Plasma temperature and hydrogen column density of M16 members}
\label{xpro_sec}

In the following sections we will address several facets of the X-ray properties of M16 members, starting from the distributions of plasma temperatures and absorption. As explained in the previous sections, values of $N_H$ and plasma temperature are associated to the X-ray source from the spectral fitting or quantile analysis. When a good fit with a thermal model is available, these values and their uncertainties are adopted from the best fit model. In the other cases, they are taken from the interpolation in the thermal grid (shown in Fig. \ref{q_image}) for the sources which lie inside the grid. For those sources without a good spectral fit with a thermal model (1T or 2T) and whose position in the diagrams in Fig. \ref{q_image} is outside the thermal grid, no values of $N_H$ and plasma temperature are provided. \par 

       \begin{figure}[]
        \centering
        \includegraphics[width=9cm]{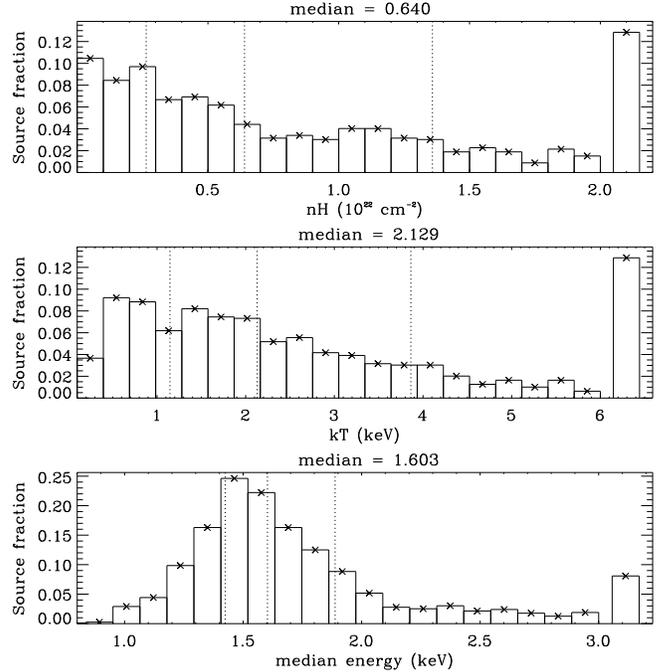}
        \caption{Distributions of the $N_H$, $kT$ and median photon energy for the X-ray sources in M16. The dotted vertical lines mark the 25\%, 50\%, and 75\% quartiles of each distribution.}
        \label{xpro_image}
        \end{figure}

 	Fig. \ref{xpro_image} shows the distributions of $N_H$, $kT$ and median photon energy of the X-ray sources in M16 (both disk-less and disk-bearing sources). The $N_H$ distribution covers a large range of values, with a median value corresponding to $A_V=3.6^m$, which is larger than the average extinction estimated by \citet{io07} from the analysis of the optical diagrams ($A_V=2.7^m$). We will show in Sect. \ref{regions_sec} that the discrepancy between these two estimates is reduced by considering the stars in the central cavity cleared by NGC~6611. Part of the large spread of the $N_H$ distribution is purely statistical, but since about 55\% of sources with lowest $N_H$ are clustered in the cavity cleared by NGC~6611 or southward (where the extinction is lower, see \citealt{io10}), while a similar fraction of the sources with highest $N_H$ mostly lie in the trunks, in the SFO30 cloud and in the north-east, part of the spread in the $N_H$ distribution is due to the strong differential reddening across the nebula. \par
The median values of both $kT$ and median photon energy are similar to those observed in other star-forming regions (for instance in Orion, \citealt{Fei02}, and in NGC~1893, \citealp{Cara11}). The two populations of sources with $kT<1.14\,keV$ and $kT>4.09\,keV$ (which are the quantiles at 80\% and 20\% of the $kT$ distribution), have different fractions of variable sources, suggesting that part of the spread observed in the $kT$ distribution can be due to X-ray variability, with some sources observed during periods of quiescent emission and others during flares. In fact, by adopting a threshold for variability of $P_{KS}<0.005$ (see Appendix \ref{cata_sec}), 42\% of the sources in the high-temperature tail of the $kT$ distribution can be variable, while among those with $kT<1.14\,keV$ only 14\% can be variable. This  suggests a larger contribution from flares in the X-ray emission of the sources with highest plasma temperature. Besides, the presence of sources with high plasma temperature ($kT>4.09$) among those with constant light curve and low luminosity supports the hypothesis that ``quiescent'' X-ray emission from magnetically active stars can arise from the merging of coronal microflares \citep{Dra00,Car07}. \par

      \begin{figure*}[]
        \centering
        \includegraphics[width=6.5cm]{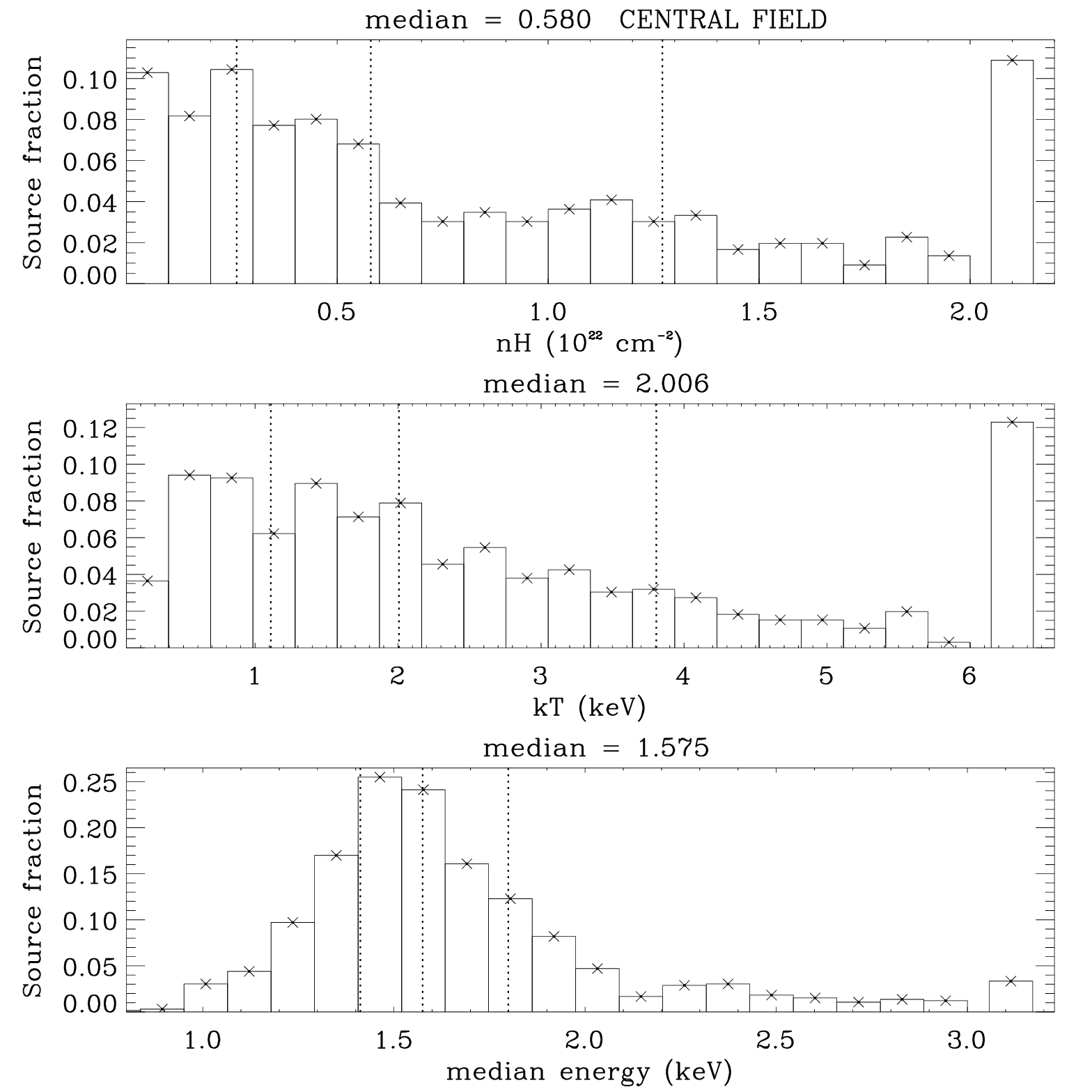}
        \includegraphics[width=6.5cm]{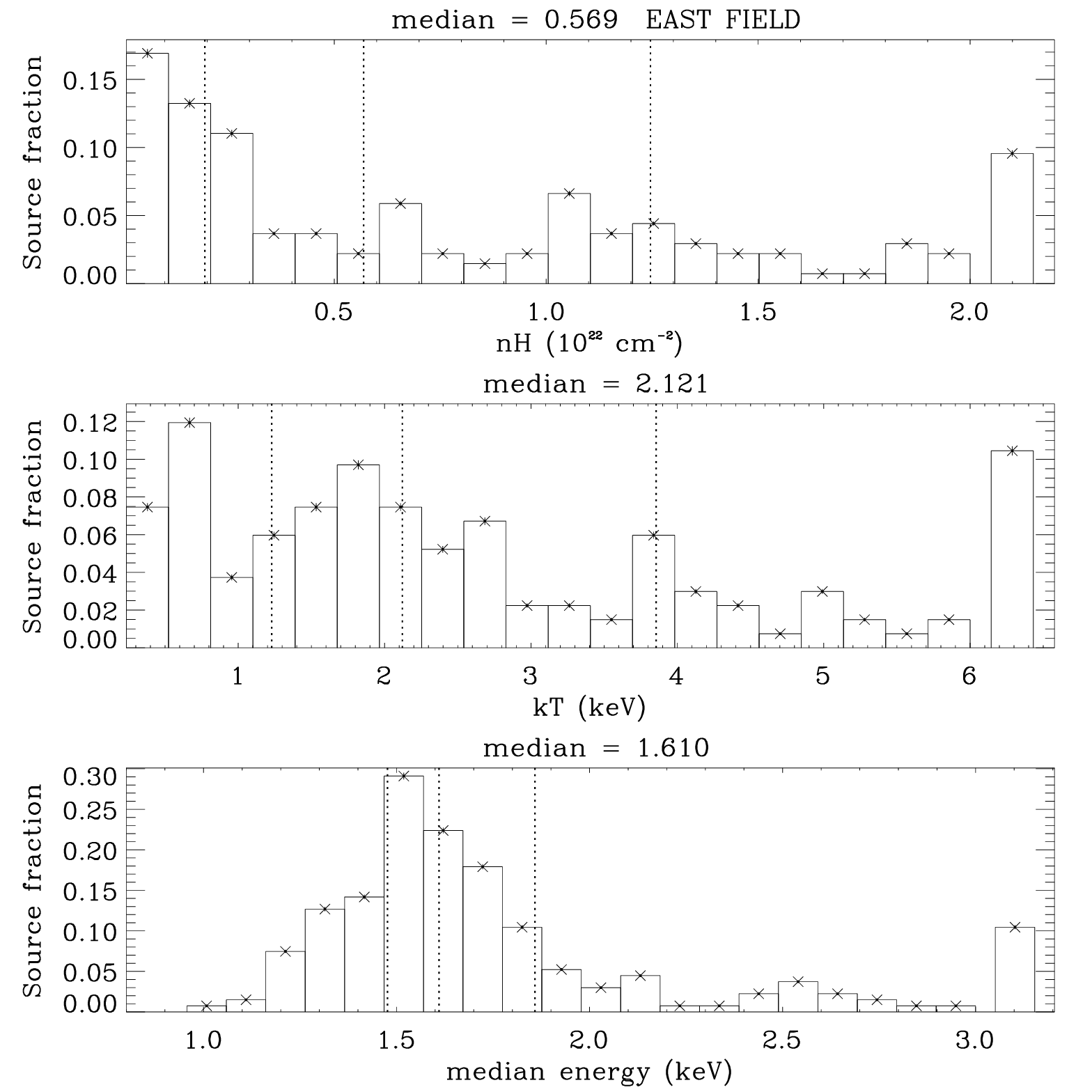}
        \includegraphics[width=6.5cm]{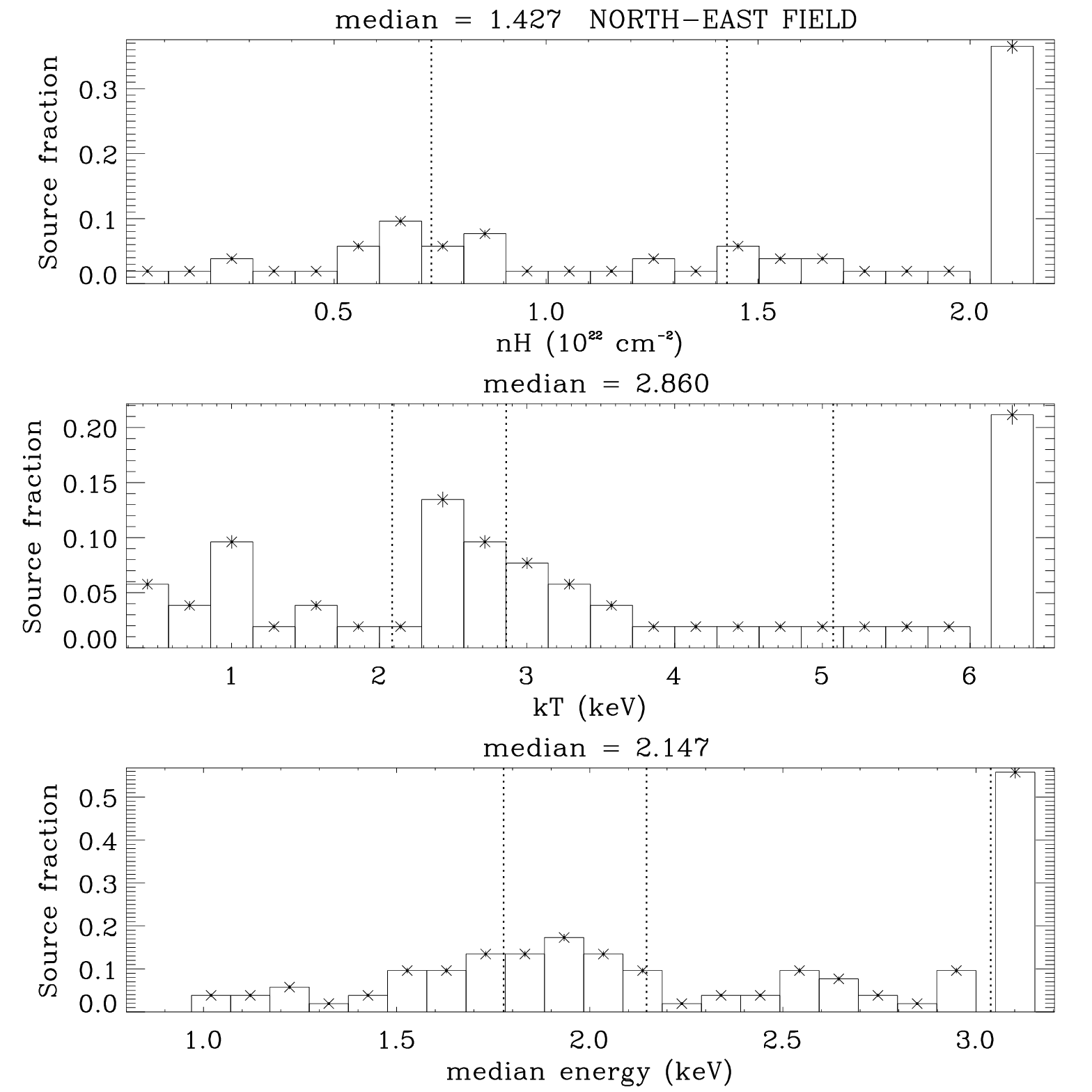}
        \caption{Distributions of the $N_H$, $kT$ and median energy of the X-ray sources in M16 falling in the three observed fields. The dotted vertical lines mark the 25\%, 50\%, and 75\% quartiles of each distribution.}
        \label{xpro_regions}
        \end{figure*}

	Fig. \ref{xpro_regions} shows the distributions of $N_H$, plasma temperature and median photon energy separately for sources in the $c-field$ (left panel), $e-field$ (central panel), and $ne-field$ (right panel). The distributions for the populations of the $c-field$ and $e-field$ are similar, reflecting the similar age and extinction in these two fields as found by \citet{io10}. The largest differences arise with the $ne-field$, where the median hydrogen column density is $N_H=1.4\times10^{22}\,cm^{-2}$, corresponding to $A_V=7.98^m$, which is higher than the extinction and median $N_H$ found in the other two fields ($3.2^m$). This mostly explains the lack of sources with soft spectra in the $ne-field$ compared with the other two fields, and the higher $kT$ median value ($2.86\,keV$ versus $2.01\,keV$ and $2.12\,keV$).	\par

\section{X-ray luminosity of M16 members}
\label{lx_sec}

	\subsection{The X-ray Luminosity Function}
	\label{xlf_sec}

	X-ray luminosities and their uncertainty have been obtained from the spectral or quantile analysis with the same criteria adopted for plasma temperature and hydrogen column (Sect. \ref{xpro_sec}). In addition, for the sources without a good spectral fit and that are outside the thermal grid in the diagrams in Fig. \ref{q_image}, they have been inferred directly from the count rates, using as conversion factor the median $L_X/(count\_rate)$ ratio found for the other sources (which ranges from $7\times10^{33}$ to $1\times10^{34}$ in the three fields). \par
The total luminosity of M16 is equal to $log\left( L_X \right)=34.90\,erg\cdot s^{-1}$, with an individual source median value of $log\left( L_X \right)=30.4^{+0.1}_{-0.6}\,erg\cdot s^{-1}$. In this estimate we excluded a candidate disk-less member with very high luminosity $log\left( L_X \right)=34.1^{+0.1}_{-0.5}\,erg\cdot s^{-1}$. This luminosity, in fact, is apparently too high to be correct, since it is two orders of magnitude higher than the $L_X$ of the second most luminous star in our sample. Its spectrum has been fitted with a 2 temperature thermal model, with a soft primary component ($kT_1=0.11 \pm 0.01\,keV$) and a harder secondary component with $kT_2=1.7 \pm 0.6\,keV$, but the emission measure of the hard component is $10^5$ times fainter than that of the soft component. One possibility is that this is a foreground source with an overestimated distance, or a background High Mass X-ray Binary, but both the $N_H$ value (equal to $4.45\times10^{22}\,cm^{-2}$) and its position in the optical color-magnitude diagrams suggest a young and embedded star lying in the BRC-SFO30. \par

        \begin{figure}[]
        \centering
        \includegraphics[width=8cm]{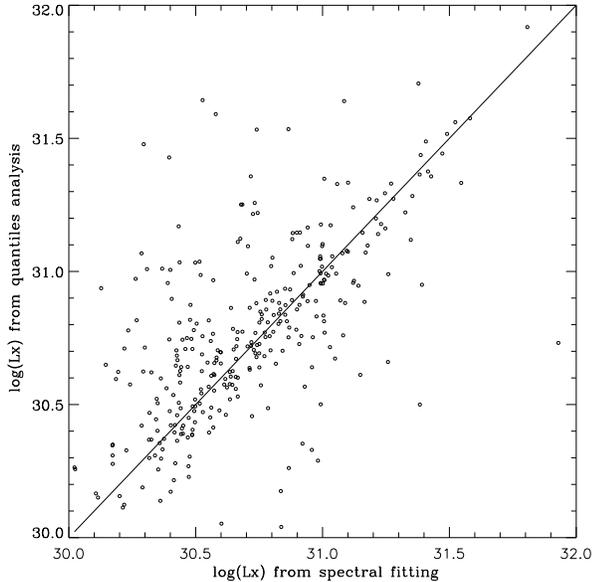}
        \caption{Values of the X-ray luminosity (in $erg\cdot s^{-1}$) calculated from the quantile analysis vs. the values obtained from spectral fitting for the X-ray sources in our sample.}
        \label{lxcompare_image}
        \end{figure}

	Fig. \ref{lxcompare_image} compares the X-ray luminosity calculated from the quantile analysis with that from the spectral fits, for the sources in our sample with a good significance and a spectral fit with good $\chi^2$. The estimate of $L_X$ from the quantile analysis agrees with that from the spectral analysis for most of the stars in Fig. \ref{lxcompare_image}, with the exception of few outliers. These outliers mostly lie in the low kT-high $N_H$ region of the grid, where the low sensitivity of the ACIS detector to the soft part of the spectrum result in large uncertainty in estimates of $N_H$ and $kT$. \par
One of the aims of this paper is to verify whether the X-ray Luminosity Function (XLF) of M16 confirms the universality of the XLFs in young clusters younger than few Myrs, as a consequence of the very slow decline of the X-ray luminosity for PMS stars younger than $10\,Myrs$ \citep{Pre05}. Several authors compared the XLFs of young clusters with that observed in the Orion Nebula Cluster (ONC) in the COUP survey, which is the most complete X-ray observation of a young cluster, complete down to $0.1\,M_{\odot}$ \citep{Get05}. A good agreement was found in IC348 and NGC~1333 ($2-3\,Myrs$, \citealp{Fei05,Wins10}), M17 ($\sim 1\,Myr$, \citealp{Broo07}), NGC~6357 ($\sim 2\,Myrs$, \citealp{Wang07}), NGC~2244 ($2-3\,Myrs$, \citealp{Wang08}), and Cyg~OB2 ($3-5\,Myrs$, \citealp{Wright10}). In this paper we adopt the same approach, comparing the XLF of M16, where the X-ray emitting population has an age ranging from $<1-3\,Myrs$ \citep{io07}, with that of the ONC. \par

        \begin{figure}[]
        \centering
        \includegraphics[width=8cm]{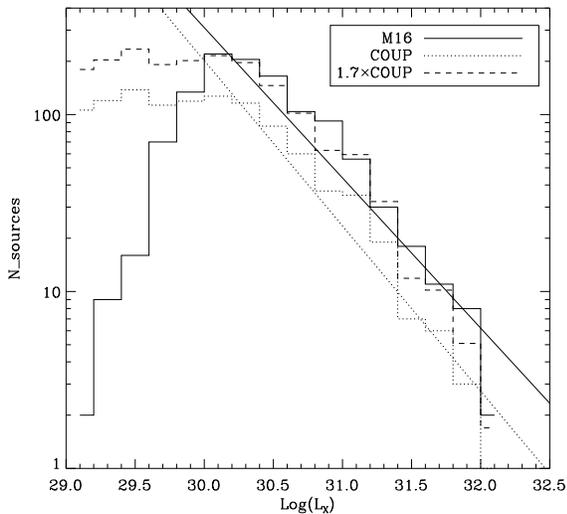}
        \caption{X-ray Luminosity Functions for M16 and Orion members. The dashed histogram is the Orion XLF scaled up to the M16 XLF. The continuous and dotted lines shows the slope of the linear fits of the XLFs for $log \left( L_X \right)> 30 \, erg \cdot s^{-1}$.}
        \label{xlf_image}
        \end{figure}

	Fig. \ref{xlf_image} shows the XLF we obtained for M16, together with that of the COUP sources in Orion. The latter is almost constant for $log \left( L_X \right)<30\,erg\cdot s^{-1}$, while the M16 XLF drops below this limit, which turns out to be the completeness limit of our observations. For luminosities higher than the completeness limit, the XLF of M16 can be fitted with a power law with index $\Gamma=-0.85 \pm 0.09$, which is consistent with that of the Orion XLF ($\Gamma=-0.93 \pm 0.08$, \citealp{Fei05}) in the same luminosity range. In this calculation, we have not considered the OB stars detected in X-rays, since the Orion OB population is significantly different from that of M16. \par
The comparison between the XLFs of Orion and M16 can help us to estimate the level of incompleteness of our sample of members down to the subsolar limit. To match the Orion XLF with the M16 XLF, we have to scale the former by 1.7. The total Orion population with $M>0.1\,M_{\odot}$ counts 1600 members \citep{Get05}. The match between the two XLFs suggest a total M16 population of 2700 members, larger than the number of members we identified in our previous work (1907 in total, incomplete a low mass regime). This is a rough estimate of the still non identified M16 population, since it depends on several properties of M16 and ONC, such as the physical extent and absorption characteristics in the respective sky areas, that are not compared properly here.\par

	\subsection{X-ray luminosity and stellar mass}
	\label{xmass_sec}

In young PMS stars, $L_X$ is generally observed to increase with the stellar mass, due to the fact that the dynamo mechanism in PMS stars is in the saturated regime, which means that it is independent of the rotational velocity and it depends only on the stellar mass and bolometric luminosity \citep{Pre05}. To study the $L_X$ vs. mass relation in M16, we estimated the masses of M16 members with optical and X-ray detections, first obtaining their intrinsic optical photometry using the 
extinction map of M16 obtained in \citet{io10}, then finding both stellar mass and age by interpolating the dereddened $V$ and $V-I$ with the colors expected from the grid of isochrones of \citet{Siess}. In this study, we focused only on the Class~III stars, since the mass estimate from the optical photometry in Class~II objects can be affected by the presence of the disk \citep{io10b}. \par

       \begin{figure}[]
        \centering
        \includegraphics[width=9cm]{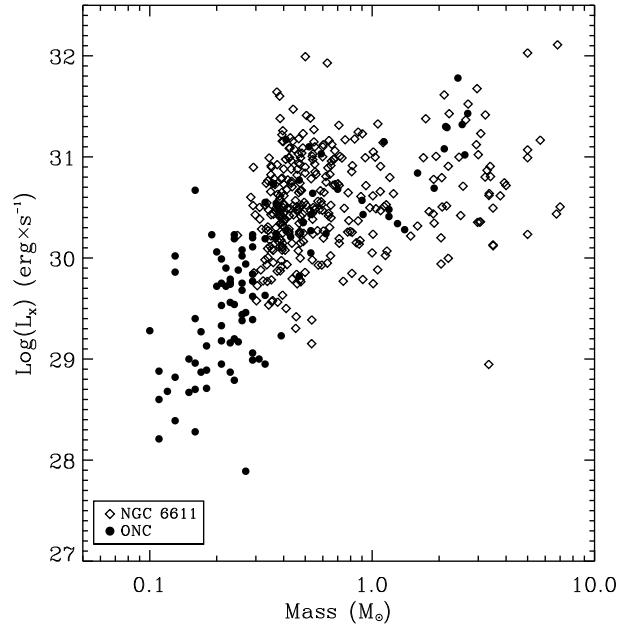}
        \caption{$log\left( L_X \right)$ vs mass for the detected Class~III members in M16 (diamonds) and Orion (dots).}
        \label{lxvsmass_figure}
        \end{figure}

	Fig. \ref{lxvsmass_figure} shows the $L_X$ vs. mass scatter plot for Class~III stars in M16 and Orion \citep{Get05}. In each mass bin, there is a large spread of $L_X$ of about $1.5-2$ orders of magnitude. This spread is mostly statistical, but also binarity can play a role \citep{Fei02}. The distribution of the low-mass stars in M16 (i.e. with $M\leq0.8\,M_{\odot}$) follows the $L_X$ vs. mass relation valid in Orion. In fact, a linear fit in this range of mass gives: 

\begin{equation}
log\left( L_X \right) = (30.9 \pm 0.1) + (1.1 \pm 0.3)\cdot log\left(M/M_{\odot} \right)
\label{lxvsmass_eq}
\end{equation}

with a slope consistent with that found in Orion \citep{Pre05}. Given the consistency of the two XLFs and assuming the universality of the IMF, the fact that the low-mass population of M16 and ONC follow the same $L_X$ vs. mass relation is not surprising. \par 
	In Fig. \ref{lxvsmass_figure} the distribution of stars more massive than $0.8\,M_{\odot}$ is totally different, being flatter (with a slope of $0.4\pm0.2$) than in the low-mass regime. Similar behavior has been observed in other young clusters. Our data do not allow us to understand the nature of the flattening of the $L_X$ vs. mass distribution observed at about $1\,M_{\odot}$. The most likely explanation is that it is due to changes in stellar interior. \citet{Fei03}, following the models of \citet{Pal93}, have shown as in the mass range between $2.4 < M/M_{\odot} < 3.9$, very young PMS stars already developed a radiative core, surrounded by a convective mantle, where deuterium burning takes place. The efficiency of the magnetic dynamo, and the X-ray activity level of the star, strongly depends on the precise boundary between these two regions, and this boundary changes with the age and accretion history of the PMS phase. For even more massive stars, the mechanism for the production of X-rays changes drastically. \par

	\subsection{X-ray activity in Class~III and Class~II members}
	\label{IIIvsII_sec}

It is still debated whether the X-ray activity in PMS stars can be affected by the presence of a circumstellar disk. Several authors have compared the X-ray luminosity of Class~III and Class~II members of young clusters, and the general picture remains unclear. In some cases, such as Chamaleon~I \citep{Fei93}, $\rho$~Ophiuchi \citep{Cas95} and IC348 \citep{Pre02}, the same level of X-ray activity in disk-less and disk-bearing members has been found; while in other clusters, such as the Taurus-Auriga region \citep{Ste01}, NGC~1893 \citep{Caramazza2012}, and NGC~2264 \citep{Fla06}, the X-ray luminosity of Class~II objects is significantly lower than that in Class~III stars. In the ONC the first attempt to find any difference in X-ray activity between these two classes of PMS stars was made by \citet{Gag95}, who found no significant difference. Later, \citet{Fla03} discovered a different emission level in accreting and non accreting stars in Orion. This result was reinforced by \citet{Pre05}, taking advantage of the completeness of the member selection with the deep COUP data. These authors found no difference in X-ray emission between stars with and without infrared excesses, detected with a diagnostic based on the $L$ photometric band, but they found a significant difference between non accreting and accreting stars, selected using the CaII line emission. Their finding suggests that the effects on the X-ray activity in disk-bearing stars are not due to the presence of the disk itself, but to active accretion. In fact, in the ONC study of \citet{Pri08}, where the disk-bearing stars have been selected by means of their infrared excesses, the level of X-ray emission in Class~II and Class~III stars is only marginally different. \par
\begin{figure*}[]
        \centering
        \includegraphics[width=8cm]{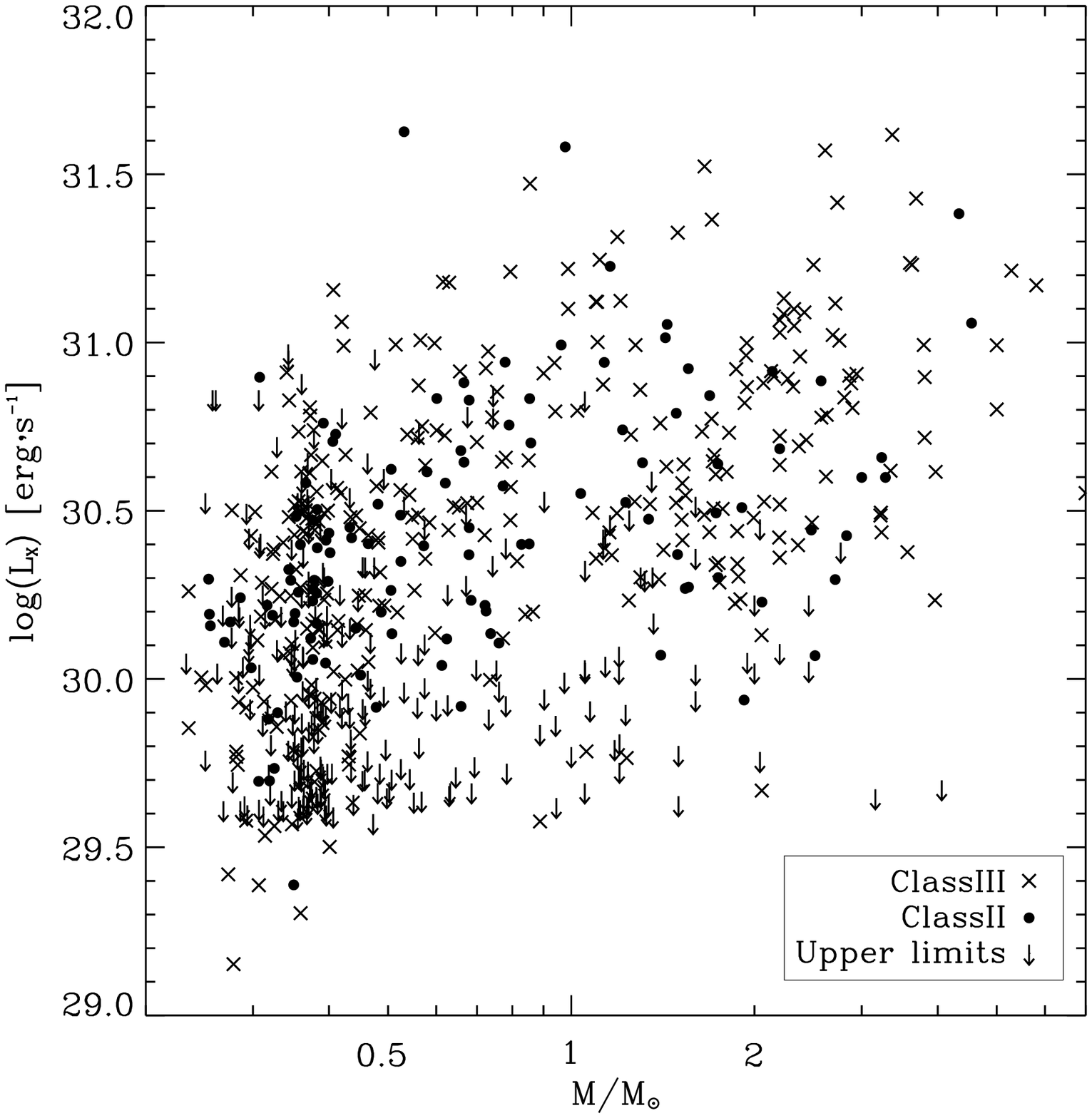}
	\includegraphics[width=8cm]{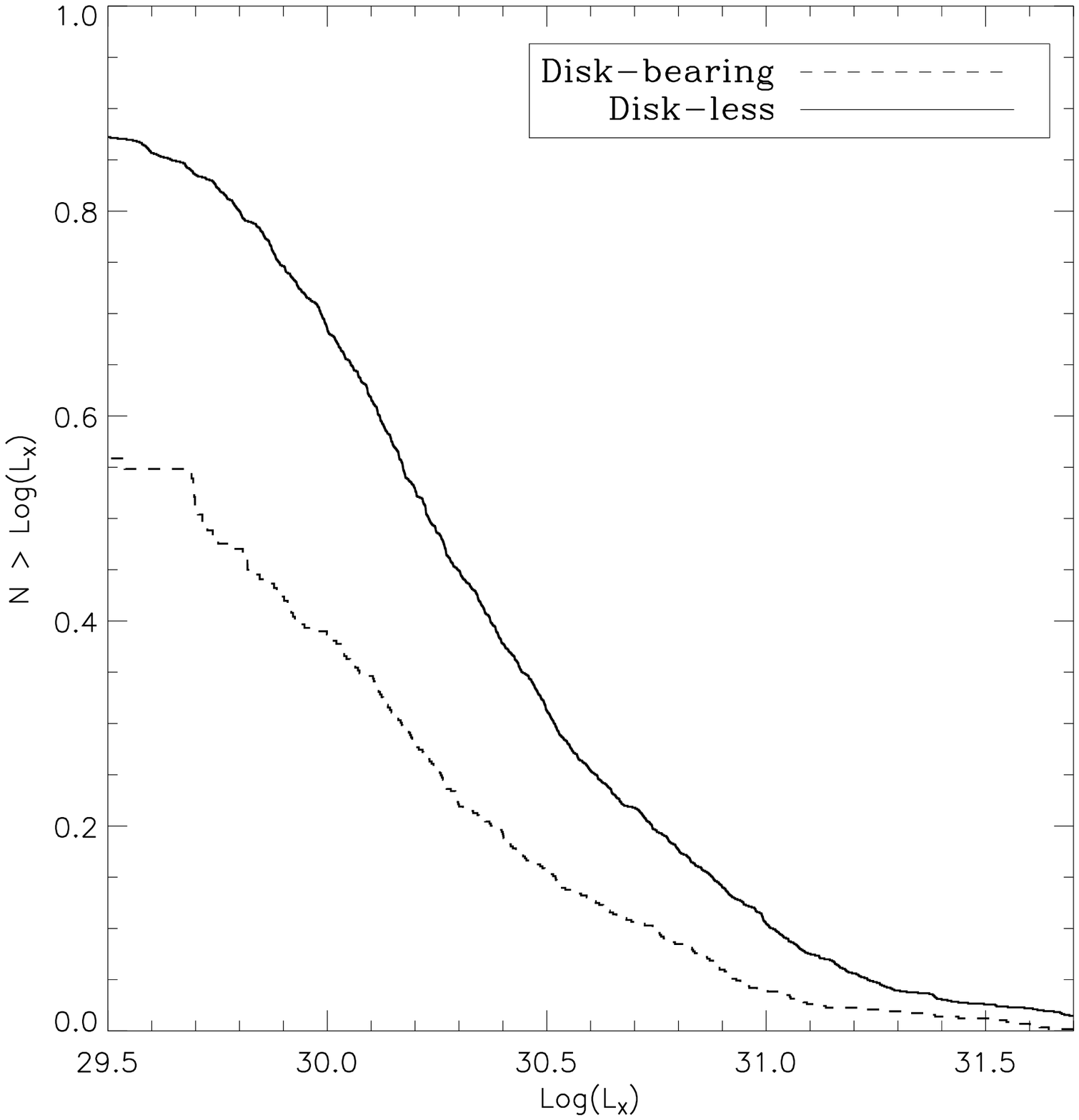}
        \caption{Left panel: $log\left( L_X \right)$ vs. $M/M_{\odot}$ for the Class~III (crosses), Class~II (filled dots) and upper limits (arrows) in M16. Right panel: X-ray luminosity function for completed Class~III and Class~II sources with mass in the $0.2 \leq M/M_{\odot} \leq 7$ range.}
        \label{IIIvsII_figure}
        \end{figure*}

	The present study of the stellar population of M16 can help to shed some light on the effects of the circumstellar disk in the X-ray emission in young stars, taking advantage of the accurate selection of both Class~II and Class~III members. However, a reliable comparison between the X-ray luminosity distributions of Class~II and Class~III objects can be attempted only if the two samples are complete. Class~II sample (219 stars) can be easily completed using the upper limits of $L_X$ of the undetected Class~II objects (see Sect. \ref{uplim_sec}). These upper limits, together with the X-ray luminosity of detected Class~II and Class~III objects, are shown in the left panel of Fig. \ref{IIIvsII_figure} as a function of stellar mass. \par

We cannot complete the Class~III sample with the same approach adopted for Class~II stars, since we have not a representative sample defined independently from X-ray information. However, we can estimate the fraction of Class~III members with X-ray emission below our completeness limit using the COUP data of Orion. Our sample of Class~III objects, as well as the Class~II, is more massive than $0.2\,M_{\odot}$ \citep{io10}, and in Orion 15\% of the Class~III objects with $M>0.2\,M_{\odot}$ have $log\left( L_X \right)<30\,erg\cdot s^{-1}$. Given the similarities between the X-ray properties of the two clusters, we can suppose that this is valid also for M16, i.e. that the undetected Class~III stars more massive than $0.2\,M_{\odot}$ are 15\% of the detections (124 sources). Regarding their upper $L_X$ limit, since our aim is to verify whether the X-ray activity in Class~III sources is higher than that in Class~II objects, we can consider the most conservative case, resulting in the lowest $L_X$ distribution for Class~III stars: that the upper limit of the X-ray luminosity of the missing Class~III stars is equal to the lowest $L_X$ measured for Class~III objects in the COUP survey \citep{Get05}, equal to $log \left( L_X \right) = 27.89\,erg/s$. We then calculated the XLFs for the ``completed'' samples of Class~III (detected Class~III sources plus 124 sources with $log \left(L_X \right)=27.89\,erg/s$) and Class~II objects (detected Class~II objects plus the upper limits of the undetected Class~II objects.), which are shown in in the right panel of Fig. \ref{IIIvsII_figure}. The emission level of the completed sample of Class~II objects is evidently lower than that of the Class~III, and we have verified with the ASURV\footnote{Astronomy Survival Analysis, http://www.astro.psu.edu/statcodes} \citep{FeiNel85} statistical package that the probability that the two XLFs are drawn from the same parent distribution is null. In the left panel, a few sources with $log \left(L_X \right)=29.5\,erg/s$ are present (fainter than our detection limit). They are low significance sources, whose luminosity has been inferred directly from their count-rates. In conclusion, in M16 the level of X-ray emission is higher in Class~III objects than in disk-bearing stars, supporting the idea that the presence of a disk results in a lower X-ray activity. We cannot discern between accreting and non accreting disks, but it can be expected that, since the average age of the M16 population is $\sim 1\,Myrs$ \citep{io07}, a large fraction of stars with disks is still actively accreting. \par
	There is more than one hypothesis to explain the observed lower X-ray emission level in Class~II than Class~III stars. The accreting material, for instance, could be responsible for a higher absorption in disk-bearing objects. In fact, the correction we made for the extinction can be not completely appropriate for disk-bearing stars. For these objects, it would be more appropriate the use of an extinction model which takes into account the presence of both interstellar and circumstellar obscuring material (Flaccomio et al., in preparation). It is also possible that the accreting material deforms the large scale stellar magnetic field \citep{Rom04} or that it increases the gas density in the magnetic loops around the stars resulting in a less efficient heating of the material by the energy released during flares \citep{Pre05}, even if this would require a fraction of the stellar surface covered by hot spots larger than the typical values observed in CTTS \citep{Muz01}. The last possibility is that the accretion phenomenon changes the internal structure of the star, as pointed by \citet{Wuc03}, who found that even with small accretion rates the central star is no longer fully convective. A different hypothesis is that the presence of the accreting inner disk is not the cause of the lower X-ray activity, but a consequence, since, as discovered by \citet{Dra09}, a high X-ray flux emitted by the central star can enhance the photoevaporation of the disk.  \par
Only one Class~I object has been detected in X-rays. Its position is $\alpha=20:33:51.237$ and $\delta=+41:25:10.68$, and it has only IRAC and Chandra detections with a high X-ray luminosity $log\left( L_X \right)=31.49\,erg\cdot s^{-1}$ derived from the spectral fit with a 1 temperature thermal model. This source lies in the embedded cluster in the $ne-field$, and it is not surprising that it has a very high extinction ($N_H=48.0\times 10^{22}\,cm^{-2}$, corresponding to $A_V=60.8^m$). The X-ray luminosity of this source is about half order of magnitude higher than that of the brightest Class~I objects in Orion \citep{Pri08}, but a similar embedded object has been identified in M17 \citep{Broo07}. However, neither the IRAC or the X-ray data allow us to discard the possibility that this source is a background AGN, whose locus in the IRAC color-color diagram (used to classify the selected disk-bearing stars) roughly coincides with that of Class~I objects. The fit of the source spectrum with a power law model is statistically acceptable and the photon index we obtained is equal to 2.03, compatible with typical values of AGN spectra. The only valid argument against the extragalactic nature of this source is its spatial correspondence with the embedded NE cluster. \par

\section{X-rays emission in high mass stars}
\label{mass_sec}

NGC~6611 is a young massive cluster comprising several massive stars. In the ACIS FoVs fall 13 O stars and 80 B stars, with one of the few O4-O5 stars known in our Galaxy with a mass of $75-80\,M_{\odot}$ (W205), and two known O binary systems (W175, an O8.5V+O5V system, \citealp{Bos99}, and W205 itself, O4V+O7.5V, \citealp{Sana09}). In our observations, we detected 85\% of the O stars (11/13), 48\% of the B0-B2 stars (24/54), and 19\% of the B3-B9 stars (7/36), among which 2 giant stars, accounting for a good sample of stars to study the mechanisms for the production of X-rays in young massive stars. In this section we use the stellar nomenclature by \citet{Win97}. \par
	The X-ray properties of the detected OB stars in NGC~6611 are summarized in Table \ref{massive_table}. For those stars observed in two observations (in all the cases in the $c-field$ and $e-field$), the table shows two values of $N_H$, $kT$, and $L_X$, corresponding to the $c-field$ and $e-field$. The variability diagnosis comes from the KS test on the light curve provided by AE (see Appendix \ref{cata_sec} for details); the spectral classification is taken from \citet{Hil93,Eva05,Duf06,Mar08}, while the binarity information is from \citet{Duch01,Herb01,Gva08,Sana09}. \par

        \begin{sidewaystable*}[]
        \centering
        \caption {X-ray properties of the OB stars.}
        \vspace{0.5cm}
        \begin{tabular}{ccccccccccc}
        \hline
        \hline
        ID & net counts &results$^*$ & $N_H\,(\times10^{22}\,cm^{-2})^{**}$ & kT$\,(keV)^{**}$ & kT$_2^{**}$ & $log\left(L_X \right)\,(erg\cdot s^{-1})^{**}$ & $log\left(L_{bol}/L_{sun} \right)$ & Variable$^{***}$ & spectral type & binary\\
        \hline
W205  &  754.7	   &F&	0.54	  &	0.63	 &     &32.17	   	&5.58      &   Y   &     O4III,O4V,O5    &       Y   \\
W175  &  1258.0    &F&	0.78	  &	0.37	 & 3.62&32.65	   	&5.49      &   Y   &     O5V+O8.5V       &       Y   \\
W197  &  1041.0    &Q&	0.05	  &	0.93	 &     &31.47	  	&5.23      &   Y   &     O7V	         &       Y   \\
W222  &  176.2	   &F&	1.69	  &	0.36	 &     &32.09	   	&5.14      &   N   &     O7V,O7III       &  	     \\
W246  &  272.2	   &F&	0.67	  &	0.53	 &     &31.70	   	&5.62      &   N   &     O7II	         &  	     \\
W401  &  413.3	   &F&	0.15	  &	0.54	 &     &31.26	   	&4.86      &   N   &     O8.5V	         &  	     \\
W161  &  208.6	   &Q&	1.24	  &	0.34	 &     &32.33	   	&4.86      &   N   &     O8.5V	         &  	 ?   \\
W166  &  151.4	   &F&	0.45	  &	0.54	 &     &31.06	   	&4.82      &   N   &     O8.5,O9V        &  	     \\
W314  &  99.6	   &F&	0.19	  &	0.62	 &     &30.64	   	&4.52      &   N   &     O9V,B0V	 &  	 Y   \\
W280  &  60.9	   &F&	0.51	  &	0.59	 &     &31.05	   	&4.68      &   Y   &     O9.5Vn	         &  	     \\
W367  &  119.6	   &F&	0.01	  &	0.44	 &     &30.59	   	&5.15      &   N   &     O9.7III,O9.5V   &  	     \\
W188  &  36.6	   &F&	0.74	  &	0.97	 &     &30.55	   	&4.26      &   N   &     B0V	     	 &	 Y   \\
W259  &  21.5	   &N&	0.14	  &	10.0	 &     &30.06	   	&4.04      &   Y   &     B0.5V	     	 &	     \\
W150  &  19.91	   &Q&	0.05	  &	1.01	 &     &29.62	   	&4.04      &   Y   &     B0.5V	     	 &	     \\
W469  &  12.6	   &Q&	1.12	  &	0.70	 &     &30.51	   	&4.04      &   ?   &     B0.5Vn	     	 &	 ?   \\
W351  &  15.7	   &Q&	0.05	  &	2.90	 &     &29.76	   	&3.82      &   N   &     B1V	     	 &	     \\
W343  &  33.6, 30.5&F&	0.09, 0.67&	22.3, 2.62&    &30.24, 30.30	&3.82      &   Y   &     B1V	     	 &	 Y   \\
W25   &  65.5	   &F&	0.83	  &	1.49	 &     &30.82	   	&3.93      &   N   &     B1V,B0.5V       &  	 Y   \\
W125  &  29.6	   &Q&	0.33	  &	1.67	 &     &30.17	   	&3.82      &   Y   &     B1.5V,B1V       &  	 Y   \\
W207  &  196.0	   &F&	0.51	  &	1.95	 &     &31.07	   	&3.82      &   Y   &     B1V	     	 &	     \\
W231  &  156.3	   &F&	0.49	  &	2.97	 &     &30.99	   	&3.82      &   Y   &     B1V	     	 &	     \\
W254  &  3.58	   &Q&	0.19	  &	0.30	 &     &29.09	   	&3.82      &   ?   &     B1V	     	 &	 Y   \\
W444  &  17.7	   &Q&	0.09	  &	2.08	 &     &30.02	   	&3.63      &   ?   &     B1V, B1.5V      &  	 ?   \\
W296  &  42.3	   &F&	0.17	  &	1.88	 &     &30.23	   	&3.57      &   N   &     B1.5V	     	 &	     \\
W601  &  31.3	   &F&	0.11	  &	1.72	 &     &30.22	   	&3.57      &   Y   &     B1.5V	     	 &	     \\
W421  &  93.5, 118.6&F&	0.48, 1.67 &	2.64, 1.15&     &30.81, 31.30	&3.57      &   N   &     B1.5:V	     	 &	     \\
W80   &  42.1	   &F&	0.09	  &	4.14	 &     &30.33	   	&3.57      &   N   &     B1V,B2V	 &  	     \\
W300  &  14.5	   &Q&	0.05	  &	1.01	 &     &29.54	   	&3.45      &   N   &     B2V,B1.5V       &  	     \\
W227  &  140.3	   &F&	1.58	  &	0.44	 &38.61&31.65	   	&3.45      &   Y   &     B2V,B1.5V       &       Y   \\
W269  &  16.7	   &Q&	1.64	  &	0.62	 &     &30.89	   	&3.45      &   N   &     B2V,B1.5V       &  	     \\
W228  &  195.8	   &F&	0.68	  &	4.70	 &     &31.21	   	&3.32      &   Y   &     B2V	     	 &	     \\
W251  &  50.45	   &F&	0.35	  &	2.17	 &     &30.44	   	&3.32      &   Y   &     B2V	     	 &	     \\
W607  &  145.9	   &Q&	0.43	  &	0.43	 &     &31.04	   	&2.81      &   N   &     B3V	     	 &	     \\
W371  &  225.6	   &F&	0.44	  &	9.05	 &     &31.23	   	&3.17      &   Y   &     B4V,B0.5V       &  	     \\
W336  &  8.7	   &Q&	0.12	  &	0.3	 &     &29.40	   	&3.32      &   ?   &     B5III	     	 &	     \\
W276  &  93.76	   &F&	0.25	  &	2.93	 &     &30.71	   	&2.14      &   N   &     B6V	     	 &	     \\
W364  &  7.6	   &Q&	0.51	  &	2.25	 &     &29.88	  	&2.05      &   ?   &     B7V	     	 &	 Y   \\
W243  &  33.6	   &F&	1.37	  &	0.84	 &     &30.74	   	&2.01      &   Y   &     B4V,B8V	 &  	 Y   \\
W221  &  64.1	   &F&	0.51	  &	2.00	 &     &30.62	   	&1.71      &   N   &     B8V	    	 &	     \\
W322  &  5.7	   &Q&	0.16	  &	0.3	 &     &29.19	   	&1.95      &   ?   &     B8V	    	 &	     \\
W400  &  83.9	   &F&	0.20	  &	1.67	 &     &30.60	   	&          &   N   &     B9III,B8IV      &  	 Y   \\
W310  &  110.5	   &F&	0.35	  &	0.69	 &     &30.84	   	&          &   N   &     B+G	         &       Y   \\
        \hline
        \hline
        \multicolumn{11}{l}{$^*$: $F$ spectral fit, $Q$ quantiles, $C$ conversion from the count rate.} \\
        \multicolumn{11}{l}{$^{**}$: Two values if the star has been observed twice.} \\
        \multicolumn{11}{l}{$^{***}$: $?$ no light curve.} \\
        \label{massive_table}
        \end{tabular}
        \end{sidewaystable*}

	\subsection{X-ray emission from the O stars}
	\label{ostars_sec}

	Generally, O and late B stars are sources of soft ($kT\sim0.5\,keV$) X-rays, thought to be produced in a myriad of small shocks in their  radiatively accelerated winds \citep{Owo99}. This emission is usually slowly variable, and, since the wind intensity scales with $L_{bol}$, X-ray emission also scales with $L_{bol}$, usually approximately as $L_X=10^{-7}\times L_{bol}=10^{31}-10^{33}\,erg\cdot s^{-1}$ \citep{Harnd79,Palla81}. However, a large scatter is usually observed, mostly due to a wide range of shock filling factors and binarity. Usually, a thermal emission spectrum observed in an OB star with $kT=0.5-0.7\,keV$ and with constant or slightly variable light curve is a signature that the X-ray emission is produced by the self-shocked wind. In some cases, however, also an intermediate/hard component in the X-ray emission from O stars has been observed: a moderate hard component ($kT=2-3\,keV$) rotationally modulated or with rapid variability \citep{Ste05}, or a very hard thermal component due to shocks in the magnetically confined wind in the stellar equatorial plane in stars with strong magnetic dipole \citep{Bab97}, wind-wind collision in close massive binaries \citep{Pol05}, which can be also source of soft x-rays emission \citep{Gagne11}, or a hard non-thermal component caused by inverse Compton scattering of UV stellar photons by relativistic charged particles in the stellar wind \citep{Che91}. \par
\begin{figure}[]
        \centering
        \includegraphics[width=9cm]{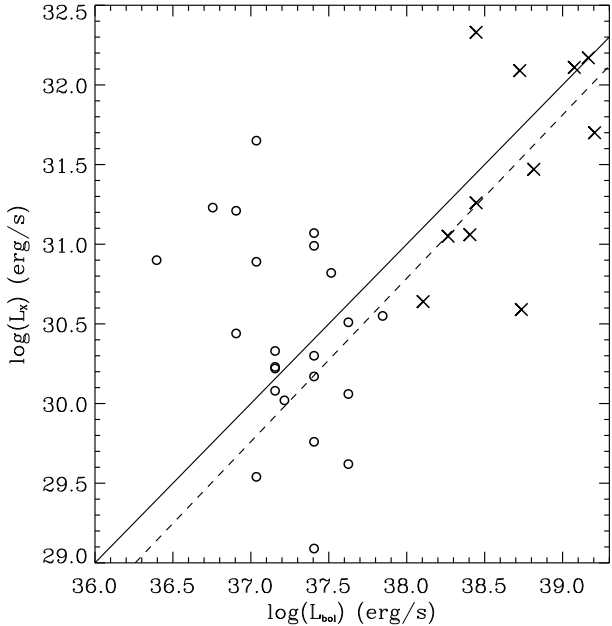}
        \caption{$log\left( L_X \right)$ vs. $log\left( L_{bol} \right)$ for the detected O (crosses) and B (circles) stars in NGC~6611. The dashed line is the best fit line calculated for for stars with $log\left( L_{bol} \right)>37\,erg\cdot s^{-1}$, the solid line is the typical $L_X\sim10^{-7}\cdot L_{bol}$ relation.}
        \label{lxvslbol_fig}
        \end{figure}

As listed in Table \ref{massive_table}, all the O stars in NGC~6611 have soft spectra (with one exception that will be discussed later), suggesting that in M16 the shocked wind emission is the only mechanism producing X-rays in the O stars. Fig. \ref{lxvslbol_fig} shows the $log\left( L_X \right)$ vs. $log\left( L_{bol} \right)$ relation for the OB stars in NGC~6611. The best linear fit for stars with $L_{bol}>10^{37}\,erg\cdot s^{-1}$ (the sample of B stars with $log\left( L_{bol} \right)<37\,erg\cdot s^{-1}$ is strongly incomplete) gives a slope of $0.7\times10^{-7}$ (dashed line), similar to the $L_X\sim10^{-7}\cdot L_{bol}$ relation (represented by the solid line) typical of the shocked wind emission, as found first by \citet{Harnd79,Palla81} and then observed in other massive star-forming regions (i.e. in the Carina region, \citealp{Naze11}). \par

	A significant level of hard emission should be expected in principle from the O4V and O7.5V binary system W205, from the collision of the intense stellar wind emitted from the two components (expected to be $\sim2.0\times10^{-6}\,M_{\odot}/year$ and $\sim0.35\times10^{-6}\,M_{\odot}/year$; \citealp{Smith2006}), but the separation between the two stars is too large (\citealp{Sana09} reported a rotation period of hundreds of years) to expect a significant wind+wind X-ray emission. \par
\begin{figure}[]
        \centering
        \includegraphics[width=9cm]{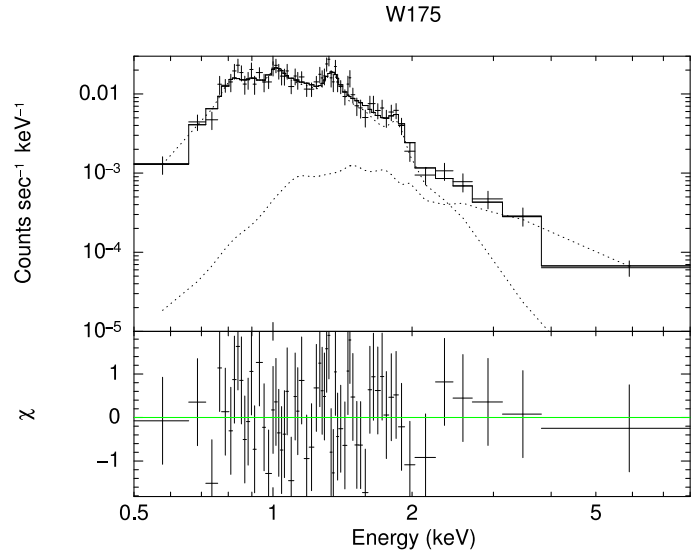}
        \caption{Thermal plasma model for W175. The two temperature components are shown (dotted lines).}
        \label{175fit_fig}
        \end{figure}

Only for W175 the spectral fit suggests the presence of a possible hard component. This system is composed of an O8.5V and an O5V stars, with a third intermediate mass (spectral type A-F) component reported by \citet{Duch01}. The separation between the two more massive components is about $1200\,AU$ \citep{Sana09}. As expected, W175 is very bright in our Chandra observations, with 1258 net counts and $log\left( L_X \right)=32.65\,erg\cdot s^{-1}$. The spectral fit with abundances fixed at standard values is very poor, with $\chi_{\nu}^2=2.06$ indicating a statistically unacceptable fit. An acceptable spectral fit ($\chi_{\nu}^2=0.86$) was instead obtained adopting a 2 temperature thermal model with abundances for Iron and Oxygen allowed to vary. The values of $N_H$, $kT$ and $L_X$ of W175 obtained with this fit are reported in Table \ref{massive_table}, and the abundances found are $O=2.3\pm 0.7\,O_{\odot}$ and $Fe=0.7\pm 0.2\,Fe_{\odot}$. In O stars a subsolar abundance of iron can be expected, as predicted, for example, by the model developed by \citet{Zhek07}, but in the case of W175 the iron abundance is only marginally subsolar. This spectral fit suggests the presence of a hard component with a temperature $kT=4\pm 2\,keV$, even if the emission measure of this hot plasma is more than 20 times smaller than that of the cold component, that can be produced in the wind-colliding zone such as observed in other similar systems (i.e. HD 93129A, \citealp{Gagne11}). Fig. \ref{175fit_fig} shows the 2T thermal plasma model fitted to W175 spectrum. \par

	\subsection{X-ray emission from B and intermediate-mass stars}
	\label{bstars_sec}

In the sample of early-B stars of NGC~6611 the ratio of bolometric flux emitted in X-ray ranges from $-5.4\leq log\left( L_X/L_{bol} \right) \leq -8.5$. This large scatter is related to the variety of mechanisms for X-ray production operating in this family of stars. Among these B stars, 10 have $kT<1\,keV$ and constant light curve, according to the KS test performed by AE. Their X-ray emission is then typical of shocked wind emission. On the other hand, 21 early B-stars have a plasma temperature $kT>1\,keV$, with six of them showing very hard spectra, with $kT>3\,keV$. For the B stars with plasma temperature $kT>1\,keV$ the KS test cannot help to discern between the possibility that the hard emission is due to an unresolved low-mass companion or to the action of one of the mechanisms described above. However, among the 6 stars with $kT>3\,keV$, five of them are not spectroscopically classified as close binary systems, and then for them the hard X-ray emission cannot be produced by colliding winds. Their light curves produced by AE suggest some flaring activity that will be studied in detail in a forthcoming paper. This emission can come from a low-mass unresolved companion in a wide orbit observed during a flaring activity, or from the B stars themselves, since, as proposed by \citet{Fei02}, magnetic reconnection events can take place on the surface of young B stars. \par
	One peculiar case is W601. This B1.5V star has been suggested to be very young ($0.016\,Myr$) and in a transition phase between the Class~II and the MS phases, with a P Cygni-like $H_{\alpha}$ line typical of the presence of outflow/infall activity \citep{Mar08}. This star has been studied in detail by \citet{Alec08}, who found an accretion rate equal to $10^{-4}\,M_{\odot}\cdot yr^{-1}$, a rotational velocity of $190\pm 10\,Km\cdot s^{-1}$, and an intense mean magnetic field equal to $3\,kG$, classifying it as a Herbig Be star. With these characteristics, it is not surprising that we detected an intermediately hard component in W601 spectrum ($kT=1.72\,keV$). The KS probability that the light curve is constant is high (0.96). \par
As expected, the fraction of detected sources in the A to late-B range of spectral classes is very low: 19\% of late-B stars and 10\% of the A-F stars, with $log\left(L_X \right)<30.4\,erg\cdot s^{-1}$, consistent with emission from a low-mass companion following the $L_X$ vs. mass relation shown in Fig. \ref{lxvsmass_figure}. One evidence that the X-ray emission detected from A-late~B stars is due to a low mass companion has been obtained by \citet{Dero11}, who found that multiplicity in a sample of X-ray detected intermediate-mass stars is about four times higher than in a control sample of stars. 
\par

\section{X-ray properties of the population of selected regions of M16}
\label{regions_sec}

       \begin{figure}[]
        \centering
        \includegraphics[width=8cm]{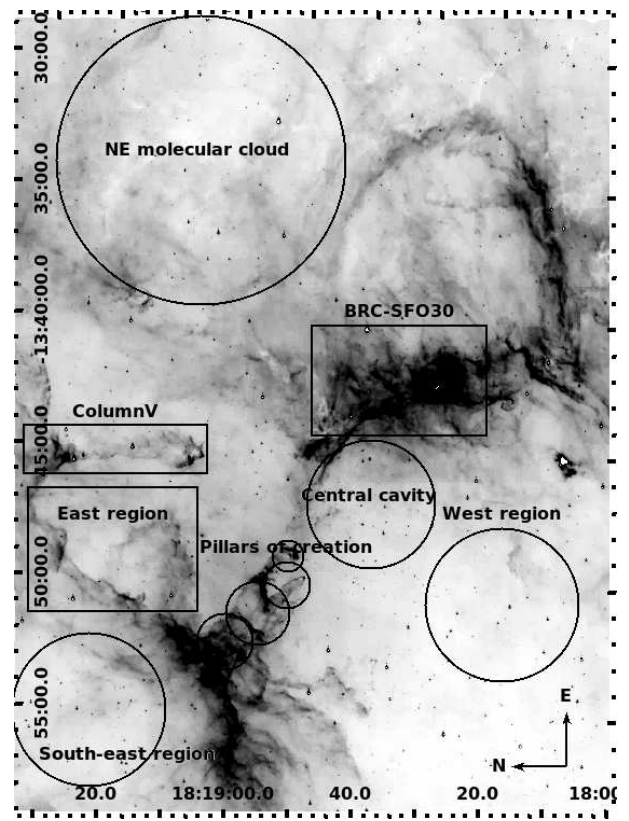}
        \caption{IRAC [8.0] image of M16 with marked all the regions whose population is analyzed in Sect. \ref{regions_sec}} 
        \label{regions_fig}
        \end{figure}

	In this appendix we analyze the X-ray properties of the YSO populations of selected regions of M16 with different characteristics. Fig. \ref{regions_fig} shows these regions on an IRAC [8.0] image. In the following discussion, the median age of the members falling in each region are taken from \citet{io10}, which also includes a detailed description of the evolutionary status of the sources falling in some of these regions. \par 
\subsection{The central cavity}
The {\it central cavity} contains the core of NGC~6611, with almost all the OB stars discussed in Sect. \ref{mass_sec} and a rich YSO population (385 members detected in X-rays, among which 21\% have a disk) with a median age of 1Myr. This region has the highest fraction of members with disks among all the regions analyzed in this section. The median $N_H$ of the X-ray members is $0.52\times 10^{22}\,cm^{-2}$, corresponding to an extinction of $A_V=2.91^m$, slightly larger than that extimated by \citet{io07} from the optical color-magnitude diagrams ($A_V=2.7^m$); the median plasma temperature is $kT=1.95\,keV$, lower than the median value of the whole population. In the {\it central cavity} there is a significant population of candidate background sources (45 sources, the 10\% of the total number of X-ray sources in this field). It is likely that most of them are really background contaminants, since we do not expect a large embedded population in this field where the molecular cloud has been cleared by the energetic radiation from the OB stars lying in this region. \par

\subsection{The pillars of creation}
	The members of the  {\it pillars of creation} are 33 (8 with disks), their age is about 2-3 Myr old in the bottom of the pillar, in the south-east, and it gets younger in the direction of the tip of the pillars. The median absorption of the X-ray sources falling in this region is typical of the cluster ($N_H=0.45\times 10^{22}\,cm^{-2}$), suggesting that almost all these sources belong to the cluster but they are not physically associated with the pillars. The median plasma temperature is $kT=2.03\,keV$, smaller than the median of the members of the whole cluster. Three X-ray sources are associated with the tip of the largest pillar, called Pillar~1, where a protostar with a mass of $4-5\,M_{\odot}$ and an intense infrared excesses has been found \citep{Mcc02}. One (at $\alpha=18:18:50.88$ and $\delta=-13:48:43.48$) has been classified as a foreground source, and its $N_H$, derived from the quantile analysis, is compatible with this classification ($N_H=2\times 10^{19}\,cm^{-2}$, corresponding to an $A_V=0.01^m$). The second source is a disk-less member located at $\alpha=18:18:49.78$ and $\delta=-13:48:56.2$. It has a spectrum fitted with a 1T thermal plasma model, with $kT=1.71\,keV$ and $N_H=0.43\times 10^{22}\,cm^{-2}$, which corresponds to the average extinction of the cluster suggesting that this source is not embedded in the pillar. The only likely X-ray counterpart of the protostar in the tip of the pillar is at $\alpha=18:18:50.31$ and $\delta=-13:48:54.31$, with 103 net counts, an absorption $N_H=3.89\times 10^{22}\,cm^{-2}$, corresponding to an $A_V=21.7^m$, and a plasma temperature of $kT=10.89\,keV$ and $log \left( L_X \right) =31.69\,erg\cdot s^{-1}$ suggesting an embedded very active young star with a X-ray thermal spectrum. The X-ray counterpart of this protostar has already been identified by \citet{Lin07}. \par

\subsection{The Column~V}
In the other pillar of M16, {\it Column~V}, fall 32 members (6 with a disk) with an age of about $1\,Myr$. Their median absorption is $N_H=0.95\times10^{22}\,cm^{-2}$, twice that of the central cavity, with the most obscured sources at the bottom of the pillar. The median value for the plasma temperature is $kT=2.18\,keV$. \citet{Mea86} discovered a bright Herbig-Haro object in the tip of Column~V. We detected one X-ray source at less than $1^{\prime \prime}$ away from it, at $\alpha=18:19:04.65$ and $\delta=-13:45:32.31$. The quantile analysis revealed a plasma temperature of $kT=3.4\,keV$, an intermediate absorption ($N_H=0.3\times10^{22}\,cm^{-2}$) and an X-ray luminosity $L_X=10^{30.1}\,erg\cdot s^{-1}$. Another candidate embedded protostar of Column~V detected in X-rays at $\alpha=18:19:07.08$ and $\delta=-13:45:22.54$ can be associated with water masers identified by \citet{Hea04}. This source is without a stellar counterparts in our catalog. It has been observed both in the $c-field$ (32.7 net counts, $log \left( L_X \right)=30.6\,erg\cdot s^{-1}$, and a spectrum fitted by a power-law) and the $e-field$ (11.8 net counts). \par

\subsection{The Bright Rimmed Cloud {\it SFO30}}
	Another region rich in cluster members is the Bright Rimmed Cloud {\it SFO30}, with 85 members (6 with a disk) detected in X-rays. In the optical and infrared diagrams, these sources are largely extincted ($A_V>5^m$). The median absorption ($N_H=1.16\times10^{22}\,cm^{-2}$) is similar to that of the sources falling in Column~V, with a median plasma temperature of $kT=2.67\,keV$. The distribution of plasma temperature of these stars, however, has a very hot tail, with the 66\% terzile at $kT=5.33\,keV$.  Considering all the X-ray sources falling in this region, there is the largest fraction of candidate members without a disk and the lowest of candidate background sources. Since, from infrared studies \citep{Ind07,io09} and detection of water masers \citep{Hea04}, a population of embedded protostars is expected to fall in this region, it is possible that the extinction here is high enough to efficiently absorb all the background sources or that these protostars are not X-ray active yet (the population of this region is on average younger than 1 Myr). \par

\subsection{The north-east embedded cluster}
The most extinguished region is the {\it NE molecular cloud}, where a young embedded cluster lies \citep{Ind07,io09}. This region has the highest fraction of X-ray sources without stellar counterparts (82 sources, 52\% of all X-ray sources falling in this region). Among the 70 candidate YSOs in this region, 12 have a disk and are classified as embedded YSOs from their IRAC colors (among them also the Class~I sources detected in X-ray discussed in Sect. \ref{lx_sec}). It is not surprising that this is the region with the highest median absorption, with $N_H=1.92\times10^{22}\,cm^{-2}$, which is four times larger than the value in the central cavity and corresponds to an $A_V=10.7^m$. \par

\subsection{The other regions}
	The remaining regions are the poorest in candidate cluster members. In the {\it East region} the 60 members detected in X-rays (12 with a disk) are older than 1 Myr, with a median absorption similar to the central cavity ($N_H=0.65\times10^{22}\,cm^{-2}$) and a median plasma temperature of $kT=2.5\,keV$. The members detected in X-ray falling in the {\it West region} (25 sources, none with a disk) have similar ages, an absorption lower than the members in the central cavity (median value $N_H=0.33\times10^{22}\,cm^{-2}$) and a median plasma temperature of $kT=2.45\,keV$. The {\it South-East region} contains the oldest X-ray detected cluster members (12 sources, one with a disk, with a median age of about $3\,Myr$), with the lowest median plasma temperature ($kT=1.72\,keV$), and a median absorption of the candidate members detected in X-rays smaller than that of the central cavity ($N_H=0.36\times10^{22}\,cm^{-2}$).

\section{Summary and conclusions}
In this paper we analyze three $80\,ksec$ Chandra/ACIS-I observations of the Eagle Nebula: one archival centered on the young open cluster NGC~6611 in the center of the nebula, one centered on the evaporating pillars named ``Column~V'', and one in the north-east region centered on an embedded cluster of very young stellar objects. The final catalog of the X-ray sources amounts to 1755 entries, classified from their photometric properties as ``{\it disk-bearing cluster members}'' (219 sources), ``{\it disk-less cluster members}'' (964), ``{\it foreground sources}'' (76), and ``{\it background sources}'' (504). \par
	Spectral properties and X-ray luminosities have been derived from spectral fitting and analysis of the photon energy quantiles. Considering all the cluster members detected in X-ray, we found a median $N_H=0.640\times 10^{22}\,cm^{-2}$, corresponding to a median extinction of $A_V=3.6^m$, and a median $kT=2.13\,keV$. The median $N_H$ is three times larger in the North-East field with respect to the central field where NGC~6611 has cleared most of the parental cloud.\par
We compared the X-ray properties of M16 members and those in the Orion Nebula Cluster, which is the best characterized sample of X-ray sources in a young cluster. The X-ray Luminosity Function has similar slope (a power law with $\Gamma=-0.85\pm0.09$) in a range of luminosities $log \left( L_X \right)>30\,erg\cdot s^{-1}$, which is the completeness limit of our observations. This allowed us to estimate that the total population of M16 should be about 2700 members. Also the slope of the $L_X$ vs. mass function is similar to that in Orion for masses smaller than $0.8\,M_{\odot}$, while for larger masses the distribution is flat around $log \left( L_X \right)=30.72\,erg\cdot s^{-1}$, with a large spread. These analogies support the universality of the X-ray luminosity functions and the $L_x$ vs. mass relation for clusters younger than $5\,Myrs$. \par
	Our study supports the evidence that the X-ray activity in disk-bearing stars is less intense than that in disk-less stars of similar mass. This result has been obtained also in other young clusters, while in some other cases no dependence with the presence of disk in the X-ray activity has been observed. The present study of M16 supports then the idea that the presence of a circumstellar disk, likely due to the accretion process, affects the coronal activity in the hosting star. \par
We took advantage of the large population of massive stars detected in X-rays (11 O stars, a 85\% detection rate, and 31 B stars, 39\%) to study the mechanism for the production of X-rays in such massive stars. All the O stars but one have soft spectra without a hard component and have constant light curves, suggesting that the X-ray photons are produced by shocks propagating in the stellar wind. A hard component has been detected only in one case (a triple system with two O stars and a A-F star), but it is only marginally significant, contributing less than the 5\% of the total spectrum. \par
	The sample of detected B stars is more heterogeneous, with 10 sources with soft spectra typical of wind emission, and 21 with $kT>1\,keV$. For these latter sources it not possible to discern the possibility that some mechanism for the production of hard X-ray photons is at work or the emission is due to unresolved companions. However, the binarity of the massive stars of M16 has been studied by several authors and most of these stars do not show evidence for such companions. In particular, among the six B stars with $kT>3\,keV$, only one is a binary system, but with 2 B stars, and all of them seem to have flares in their light curves. This result suggests that that in these young massive stars some magnetic reconnection events can occur on the stellar surface. \par   
Finally, the X-ray properties of the sources in relevant regions of M16, such as the central cavity and the Pillars of Creation, have been discussed to analyze the different properties of the M16 population in different locations of the nebula.


\acknowledgments
We thank the anonymous referee for his useful comments and suggestions. This work is based on observations performed with Chandra/ACIS-I. M.\,G.\,G. was supported by Chandra Grant GO0-11040X. J.\,J.\,D. was supported by NASA contract NAS8-39073 to the Chandra X-ray Center and thanks the Director, H.~Tananbaum, for continuing advice and support. G.\,M., S.\,S., M.\,C., and L.\,P. were supported by the contract PRIN-INAF 2009.


\appendix
\section{Source catalog}
\label{cata_sec}

	The information about the X-ray sources of M16 analyzed in this work are available in an on line catalog. The catalog is not shown here because of the large number of columns. In the following we describe the data available in the catalog. \par

\begin{itemize}
\item $RA$, $DEC$: source position in J2000.
\item $\sigma_{pos}$: the uncertainty on source position found by AE.
\item $\theta$: the source off-axis angle.
\item $PBS$: the probability associated with the null-hypothesis that there is not a real source associated with the position. Only 21 sources have the $PBS$ value larger than the threshold of 0.1 typically used with AE.
\item $Sgnf$: photometric significance computed as the ratio between the net counts in $0.5-8\,keV$ on the upper error of net counts. 
\item $Status$: class of the source, based on its optical/infrared properties (disk-bearing members, disk-less members, foreground source, background source).
\item $Cnts$: the total number of counts associated with the source in the whole energy band ($0.5-8\,keV$).
\item $Bkg\_cnts$: the observed background counts associated with the source in the whole energy band ($0.5-8\,keV$).
\item $Net\_cnts$: the net counts associated with the source in the whole energy band ($0.5-8\,keV$).
\item $PSF\_frac$:  the fraction of the PSF at 1.497 keV corresponding to the extraction region.
\item $medE$: the median photon energy in keV.
\item $N_H$: the value of the hydrogen column density in the direction of the source, in units of $10^{22}\,cm^{-2}$. This value is estimated either from the spectral fitting or quantile analysis as described in Sect. \ref{xpro_sec}, with errors.
\item $kT$: temperature of the emitting plasma, in keV. This value is estimated either from the spectral fitting or quantile analysis as described in Sect. \ref{xpro_sec}, with errors. 
\item $kT_2$: second temperature of the emitting plasma, in keV, for the sources whose spectra are well-fitted by a thermal 2-temperature plasma model, with errors.
\item $P_0$: index of the power law for the sources whose spectrum is best fitted by a power-law model.
\item $L_X$: the adopted X-ray luminosity for the source. See sect. \ref{lx_sec} for details on how $L_X$ is determined, with errors.
\item $Model$: the model used to fit the source spectrum (thermal, 2T thermal, power law). See Sect. \ref{spectra_sec}.
\item $P_{KS}$: probability based on the K-S statistic with the null hypothesis that the source is not variable (in the total band). The classification suggested by the AE manual is: no evidence for variability ($0.05<P_{KS}$); possibly variable ($0.005<P_{KS}<0.05$); definitely variable ($P_{KS}<0.005$).
\end{itemize}

All the information, but the position, $PBS$ and $Status$, are provided for all the three observations, sorted in alphabetical order ($c-field$, $e-field$, and $ne-field$), for a total of 63 columns.

\newpage
\addcontentsline{toc}{section}{\bf Bibliografia}
\bibliographystyle{aa}
\bibliography{biblio}

\begin{thebibliography}{78}
\expandafter\ifx\csname natexlab\endcsname\relax\def\natexlab#1{#1}\fi

\bibitem[{{Alecian} {et~al.}(2008){Alecian}, {Wade}, {Catala}, {Bagnulo},
  {Boehm}, {Bohlender}, {Bouret}, {Donati}, {Folsom}, {Grunhut}, \&
  {Landstreet}}]{Alec08}
{Alecian}, E., {Wade}, G.~A., {Catala}, C., {et~al.} 2008, \aap, 481, L99

\bibitem[{{Babel} \& {Montmerle}(1997)}]{Bab97}
{Babel}, J. \& {Montmerle}, T. 1997, \apjl, 485, L29+

\bibitem[{{Bosch} {et~al.}(1999){Bosch}, {Morrell}, \& {Niemela}}]{Bos99}
{Bosch}, G.~L., {Morrell}, N.~I., \& {Niemela}, V.~S. 1999, RMxAA, 35, 85

\bibitem[{{Brandt} {et~al.}(2001){Brandt}, {Alexander}, {Hornschemeier},
  {Garmire}, {Schneider}, {Barger}, {Bauer}, {Broos}, {Cowie}, {Townsley},
  {Burrows}, {Chartas}, {Feigelson}, {Griffiths}, {Nousek}, \&
  {Sargent}}]{Bra01}
{Brandt}, W.~N., {Alexander}, D.~M., {Hornschemeier}, A.~E., {et~al.} 2001,
  \aj, 122, 2810

\bibitem[{{Broos} {et~al.}(2007){Broos}, {Feigelson}, {Townsley}, {Getman},
  {Wang}, {Garmire}, {Jiang}, \& {Tsuboi}}]{Broo07}
{Broos}, P.~S., {Feigelson}, E.~D., {Townsley}, L.~K., {et~al.} 2007, \apjs,
  169, 353

\bibitem[{{Broos} {et~al.}(2010){Broos}, {Townsley}, {Feigelson}, {Getman},
  {Bauer}, \& {Garmire}}]{Bro10}
{Broos}, P.~S., {Townsley}, L.~K., {Feigelson}, E.~D., {et~al.} 2010, \apj,
  714, 1582

\bibitem[{{Caramazza} {et~al.}(2007){Caramazza}, {Flaccomio}, {Micela},
  {Reale}, {Wolk}, \& {Feigelson}}]{Car07}
{Caramazza}, M., {Flaccomio}, E., {Micela}, G., {et~al.} 2007, \aap, 471, 645

\bibitem[{{Caramazza} {et~al.}(2011){Caramazza}, {Micela}, {Prisinzano},
  {Sciortino}, {Damiani}, {Favata}, {Stauffer}, {Vallenari}, \&
  {Wolk}}]{Cara11}
{Caramazza}, M., {Micela}, G., {Prisinzano}, L., {et~al.} 2011, ArXiv e-prints

\bibitem[{{Caramazza} {et~al.}(2012){Caramazza}, {Micela}, {Prisinzano},
  {Sciortino}, {Damiani}, {Favata}, {Stauffer}, {Vallenari}, \&
  {Wolk}}]{Caramazza2012}
{Caramazza}, M., {Micela}, G., {Prisinzano}, L., {et~al.} 2012, \aap, 539, A74

\bibitem[{{Casanova} {et~al.}(1995){Casanova}, {Montmerle}, {Feigelson}, \&
  {Andre}}]{Cas95}
{Casanova}, S., {Montmerle}, T., {Feigelson}, E.~D., \& {Andre}, P. 1995, \apj,
  439, 752

\bibitem[{{Chen} \& {White}(1991)}]{Che91}
{Chen}, W. \& {White}, R.~L. 1991, \apj, 366, 512

\bibitem[{{Damiani} {et~al.}(1997){Damiani}, {Maggio}, {Micela}, \&
  {Sciortino}}]{Dami97}
{Damiani}, F., {Maggio}, A., {Micela}, G., \& {Sciortino}, S. 1997, \apj, 483,
  350

\bibitem[{{De Rosa} {et~al.}(2011){De Rosa}, {Bulger}, {Patience}, {Leland},
  {Macintosh}, {Schneider}, {Song}, {Marois}, {Graham}, {Bessell}, \&
  {Doyon}}]{Dero11}
{De Rosa}, R.~J., {Bulger}, J., {Patience}, J., {et~al.} 2011, \mnras, 415, 854

\bibitem[{{de Winter} {et~al.}(1997){de Winter}, {Koulis}, {The}, {van den
  Ancker}, {Perez}, \& {Bibo}}]{Win97}
{de Winter}, D., {Koulis}, C., {The}, P.~S., {et~al.} 1997, \aaps, 121, 223

\bibitem[{{Drake} {et~al.}(2009){Drake}, {Wright}, \& {The Chandra Cyg Ob2
  Team}}]{Dra09}
{Drake}, J., {Wright}, N., \& {The Chandra Cyg Ob2 Team}. 2009, in Chandra's
  First Decade of Discovery, ed. {S.~Wolk, A.~Fruscione, \& D.~Swartz}

\bibitem[{{Drake} {et~al.}(2000){Drake}, {Peres}, {Orlando}, {Laming}, \&
  {Maggio}}]{Dra00}
{Drake}, J.~J., {Peres}, G., {Orlando}, S., {Laming}, J.~M., \& {Maggio}, A.
  2000, \apj, 545, 1074

\bibitem[{{Duch{\^e}ne} {et~al.}(2001){Duch{\^e}ne}, {Simon}, {Eisl{\"o}ffel},
  \& {Bouvier}}]{Duch01}
{Duch{\^e}ne}, G., {Simon}, T., {Eisl{\"o}ffel}, J., \& {Bouvier}, J. 2001,
  \aap, 379, 147

\bibitem[{{Dufton} {et~al.}(2006){Dufton}, {Smartt}, {Lee}, {Ryans}, {Hunter},
  {Evans}, {Herrero}, {Trundle}, {Lennon}, {Irwin}, \& {Kaufer}}]{Duf06}
{Dufton}, P.~L., {Smartt}, S.~J., {Lee}, J.~K., {et~al.} 2006, \aap, 457, 265

\bibitem[{{Evans} {et~al.}(2005){Evans}, {Smartt}, {Lee}, {Lennon}, {Kaufer},
  {Dufton}, {Trundle}, {Herrero}, {Sim{\'o}n-D{\'{\i}}az}, {de Koter},
  {Hamann}, {Hendry}, {Hunter}, {Irwin}, {Korn}, {Kudritzki}, {Langer},
  {Mokiem}, {Najarro}, {Pauldrach}, {Przybilla}, {Puls}, {Ryans}, {Urbaneja},
  {Venn}, \& {Villamariz}}]{Eva05}
{Evans}, C.~J., {Smartt}, S.~J., {Lee}, J.-K., {et~al.} 2005, \aap, 437, 467

\bibitem[{{Feigelson} {et~al.}(2002){Feigelson}, {Broos}, {Gaffney}, {Garmire},
  {Hillenbrand}, {Pravdo}, {Townsley}, \& {Tsuboi}}]{Fei02}
{Feigelson}, E.~D., {Broos}, P., {Gaffney}, III, J.~A., {et~al.} 2002, \apj,
  574, 258

\bibitem[{{Feigelson} {et~al.}(1993){Feigelson}, {Casanova}, {Montmerle}, \&
  {Guibert}}]{Fei93}
{Feigelson}, E.~D., {Casanova}, S., {Montmerle}, T., \& {Guibert}, J. 1993,
  \apj, 416, 623

\bibitem[{{Feigelson} \& {Decampli}(1981)}]{Fei81}
{Feigelson}, E.~D. \& {Decampli}, W.~M. 1981, \apjl, 243, L89

\bibitem[{{Feigelson} {et~al.}(2003){Feigelson}, {Gaffney}, {Garmire},
  {Hillenbrand}, \& {Townsley}}]{Fei03}
{Feigelson}, E.~D., {Gaffney}, III, J.~A., {Garmire}, G., {Hillenbrand}, L.~A.,
  \& {Townsley}, L. 2003, \apj, 584, 911

\bibitem[{{Feigelson} {et~al.}(2005){Feigelson}, {Getman}, {Townsley},
  {Garmire}, {Preibisch}, {Grosso}, {Montmerle}, {Muench}, \&
  {McCaughrean}}]{Fei05}
{Feigelson}, E.~D., {Getman}, K., {Townsley}, L., {et~al.} 2005, \apjs, 160,
  379

\bibitem[{{Feigelson} \& {Nelson}(1985)}]{FeiNel85}
{Feigelson}, E.~D. \& {Nelson}, P.~I. 1985, \apj, 293, 192

\bibitem[{{Flaccomio} {et~al.}(2003){Flaccomio}, {Damiani}, {Micela},
  {Sciortino}, {Harnden}, {Murray}, \& {Wolk}}]{Fla03}
{Flaccomio}, E., {Damiani}, F., {Micela}, G., {et~al.} 2003, \apj, 582, 398

\bibitem[{{Flaccomio} {et~al.}(2006){Flaccomio}, {Micela}, \&
  {Sciortino}}]{Fla06}
{Flaccomio}, E., {Micela}, G., \& {Sciortino}, S. 2006, \aap, 455, 903

\bibitem[{{Flagey} {et~al.}(2011){Flagey}, {Boulanger}, {Noriega-Crespo},
  {Paladini}, {Montmerle}, {Carey}, {Gagn{\'e}}, \& {Shenoy}}]{Flag11}
{Flagey}, N., {Boulanger}, F., {Noriega-Crespo}, A., {et~al.} 2011, \aap, 531,
  A51+

\bibitem[{{Gagne} {et~al.}(1995){Gagne}, {Caillault}, \& {Stauffer}}]{Gag95}
{Gagne}, M., {Caillault}, J.-P., \& {Stauffer}, J.~R. 1995, \apj, 445, 280

\bibitem[{{Gagn{\'e}} {et~al.}(2011){Gagn{\'e}}, {Fehon}, {Savoy}, {Cohen},
  {Townsley}, {Broos}, {Povich}, {Corcoran}, {Walborn}, {Remage Evans},
  {Moffat}, {Naz{\'e}}, \& {Oskinova}}]{Gagne11}
{Gagn{\'e}}, M., {Fehon}, G., {Savoy}, M.~R., {et~al.} 2011, \apjs, 194, 5

\bibitem[{{Getman} {et~al.}(2005){Getman}, {Flaccomio}, {Broos}, {Grosso},
  {Tsujimoto}, {Townsley}, {Garmire}, {Kastner}, {Li}, {Harnden}, {Wolk},
  {Murray}, {Lada}, {Muench}, {McCaughrean}, {Meeus}, {Damiani}, {Micela},
  {Sciortino}, {Bally}, {Hillenbrand}, {Herbst}, {Preibisch}, \&
  {Feigelson}}]{Get05}
{Getman}, K.~V., {Flaccomio}, E., {Broos}, P.~S., {et~al.} 2005, \apjs, 160,
  319

\bibitem[{{Guarcello} {et~al.}(2010{\natexlab{a}}){Guarcello}, {Damiani},
  {Micela}, {Peres}, {Prisinzano}, \& {Sciortino}}]{io10b}
{Guarcello}, M.~G., {Damiani}, F., {Micela}, G., {et~al.} 2010{\natexlab{a}},
  \aap, 521, A18+

\bibitem[{{Guarcello} {et~al.}(2009){Guarcello}, {Micela}, {Damiani}, {Peres},
  {Prisinzano}, \& {Sciortino}}]{io09}
{Guarcello}, M.~G., {Micela}, G., {Damiani}, F., {et~al.} 2009, \aap, 496, 453

\bibitem[{{Guarcello} {et~al.}(2010{\natexlab{b}}){Guarcello}, {Micela},
  {Peres}, {Prisinzano}, \& {Sciortino}}]{io10}
{Guarcello}, M.~G., {Micela}, G., {Peres}, G., {Prisinzano}, L., \&
  {Sciortino}, S. 2010{\natexlab{b}}, \aap, 521, A61+

\bibitem[{{Guarcello} {et~al.}(2007){Guarcello}, {Prisinzano}, {Micela},
  {Damiani}, {Peres}, \& {Sciortino}}]{io07}
{Guarcello}, M.~G., {Prisinzano}, L., {Micela}, G., {et~al.} 2007, \aap, 462,
  245

\bibitem[{{Gvaramadze} \& {Bomans}(2008)}]{Gva08}
{Gvaramadze}, V.~V. \& {Bomans}, D.~J. 2008, \aap, 490, 1071

\bibitem[{{Harnden} {et~al.}(1979){Harnden}, {Branduardi}, {Gorenstein},
  {Grindlay}, {Rosner}, {Topka}, {Elvis}, {Pye}, \& {Vaiana}}]{Harnd79}
{Harnden}, Jr., F.~R., {Branduardi}, G., {Gorenstein}, P., {et~al.} 1979,
  \apjl, 234, L51

\bibitem[{{Healy} {et~al.}(2004){Healy}, {Hester}, \& {Claussen}}]{Hea04}
{Healy}, K.~R., {Hester}, J.~J., \& {Claussen}, M.~J. 2004, \apj, 610, 835

\bibitem[{{Herbig} \& {Dahm}(2001)}]{Herb01}
{Herbig}, G.~H. \& {Dahm}, S.~E. 2001, \pasp, 113, 195

\bibitem[{{Hester} {et~al.}(1996){Hester}, {Scowen}, {Sankrit}, {Lauer},
  {Ajhar}, {Baum}, {Code}, {Currie}, {Danielson}, {Ewald}, {Faber},
  {Grillmair}, {Groth}, {Holtzman}, {Hunter}, {Kristian}, {Light}, {Lynds},
  {Monet}, {O'Neil}, {Shaya}, {Seidelmann}, \& {Westphal}}]{Hes96}
{Hester}, J.~J., {Scowen}, P.~A., {Sankrit}, R., {et~al.} 1996, \aj, 111, 2349

\bibitem[{{Hillenbrand} {et~al.}(1993){Hillenbrand}, {Massey}, {Strom}, \&
  {Merrill}}]{Hil93}
{Hillenbrand}, L.~A., {Massey}, P., {Strom}, S.~E., \& {Merrill}, K.~M. 1993,
  \aj, 106, 1906

\bibitem[{{Hong} {et~al.}(2004){Hong}, {Schlegel}, \& {Grindlay}}]{Hon04}
{Hong}, J., {Schlegel}, E.~M., \& {Grindlay}, J.~E. 2004, \apj, 614, 508

\bibitem[{{Indebetouw} {et~al.}(2007){Indebetouw}, {Robitaille}, {Whitney},
  {Churchwell}, {Babler}, {Meade}, {Watson}, \& {Wolfire}}]{Ind07}
{Indebetouw}, R., {Robitaille}, T.~P., {Whitney}, B.~A., {et~al.} 2007, \apj,
  666, 321

\bibitem[{{Jansen} {et~al.}(2001){Jansen}, {Lumb}, {Altieri}, {Clavel}, {Ehle},
  {Erd}, {Gabriel}, {Guainazzi}, {Gondoin}, {Much}, {Munoz}, {Santos},
  {Schartel}, {Texier}, \& {Vacanti}}]{Jans01}
{Jansen}, F., {Lumb}, D., {Altieri}, B., {et~al.} 2001, \aap, 365, L1

\bibitem[{{Linsky} {et~al.}(2007){Linsky}, {Gagn{\'e}}, {Mytyk}, {McCaughrean},
  \& {Andersen}}]{Lin07}
{Linsky}, J.~L., {Gagn{\'e}}, M., {Mytyk}, A., {McCaughrean}, M., \&
  {Andersen}, M. 2007, \apj, 654, 347

\bibitem[{{Maggio} {et~al.}(2007){Maggio}, {Flaccomio}, {Favata}, {Micela},
  {Sciortino}, {Feigelson}, \& {Getman}}]{Mag07}
{Maggio}, A., {Flaccomio}, E., {Favata}, F., {et~al.} 2007, \apj, 660, 1462

\bibitem[{{Martayan} {et~al.}(2008){Martayan}, {Floquet}, {Hubert}, {Neiner},
  {Fr{\'e}mat}, {Baade}, \& {Fabregat}}]{Mar08}
{Martayan}, C., {Floquet}, M., {Hubert}, A.~M., {et~al.} 2008, \aap, 489, 459

\bibitem[{{McCaughrean} \& {Andersen}(2002)}]{Mcc02}
{McCaughrean}, M.~J. \& {Andersen}, M. 2002, \aap, 389, 513

\bibitem[{{Meaburn} \& {Walsh}(1986)}]{Mea86}
{Meaburn}, J. \& {Walsh}, J.~R. 1986, \mnras, 220, 745

\bibitem[{{Montmerle} {et~al.}(1983){Montmerle}, {Koch-Miramond}, {Falgarone},
  \& {Grindlay}}]{Mon83}
{Montmerle}, T., {Koch-Miramond}, L., {Falgarone}, E., \& {Grindlay}, J.~E.
  1983, \apj, 269, 182

\bibitem[{{Morrison} \& {McCammon}(1983)}]{Mor83}
{Morrison}, R. \& {McCammon}, D. 1983, \apj, 270, 119

\bibitem[{{Muzerolle} {et~al.}(2001){Muzerolle}, {Calvet}, \&
  {Hartmann}}]{Muz01}
{Muzerolle}, J., {Calvet}, N., \& {Hartmann}, L. 2001, \apj, 550, 944

\bibitem[{{Naz{\'e}} {et~al.}(2011){Naz{\'e}}, {Broos}, {Oskinova}, {Townsley},
  {Cohen}, {Corcoran}, {Evans}, {Gagn{\'e}}, {Moffat}, {Pittard}, {Rauw},
  {ud-Doula}, \& {Walborn}}]{Naze11}
{Naz{\'e}}, Y., {Broos}, P.~S., {Oskinova}, L., {et~al.} 2011, \apjs, 194, 7

\bibitem[{{Owocki} \& {Cohen}(1999)}]{Owo99}
{Owocki}, S.~P. \& {Cohen}, D.~H. 1999, \apj, 520, 833

\bibitem[{{Palla} \& {Stahler}(1993)}]{Pal93}
{Palla}, F. \& {Stahler}, S.~W. 1993, \apj, 418, 414

\bibitem[{{Pallavicini} {et~al.}(1981){Pallavicini}, {Golub}, {Rosner},
  {Vaiana}, {Ayres}, \& {Linsky}}]{Palla81}
{Pallavicini}, R., {Golub}, L., {Rosner}, R., {et~al.} 1981, \apj, 248, 279

\bibitem[{{Pollock} {et~al.}(2005){Pollock}, {Corcoran}, {Stevens}, \&
  {Williams}}]{Pol05}
{Pollock}, A.~M.~T., {Corcoran}, M.~F., {Stevens}, I.~R., \& {Williams}, P.~M.
  2005, \apj, 629, 482

\bibitem[{{Preibisch} \& {Feigelson}(2005)}]{Pre05}
{Preibisch}, T. \& {Feigelson}, E.~D. 2005, \apjs, 160, 390

\bibitem[{{Preibisch} \& {Zinnecker}(2002)}]{Pre02}
{Preibisch}, T. \& {Zinnecker}, H. 2002, \aj, 123, 1613

\bibitem[{{Prisinzano} {et~al.}(2008){Prisinzano}, {Micela}, {Flaccomio},
  {Stauffer}, {Megeath}, {Rebull}, {Robberto}, {Smith}, {Feigelson}, {Grosso},
  \& {Wolk}}]{Pri08}
{Prisinzano}, L., {Micela}, G., {Flaccomio}, E., {et~al.} 2008, \apj, 677, 401

\bibitem[{{Puccetti} {et~al.}(2006){Puccetti}, {Fiore}, {D'Elia}, {Pillitteri},
  {Feruglio}, {Grazian}, {Brusa}, {Ciliegi}, {Comastri}, {Gruppioni},
  {Mignoli}, {Vignali}, {Zamorani}, {La Franca}, {Sacchi}, {Franceschini},
  {Berta}, {Buttery}, \& {Dias}}]{Puccetti2006}
{Puccetti}, S., {Fiore}, F., {D'Elia}, V., {et~al.} 2006, \aap, 457, 501

\bibitem[{{Romanova} {et~al.}(2004){Romanova}, {Ustyugova}, {Koldoba}, \&
  {Lovelace}}]{Rom04}
{Romanova}, M.~M., {Ustyugova}, G.~V., {Koldoba}, A.~V., \& {Lovelace},
  R.~V.~E. 2004, \apjl, 616, L151

\bibitem[{{Sana} {et~al.}(2009){Sana}, {Gosset}, \& {Evans}}]{Sana09}
{Sana}, H., {Gosset}, E., \& {Evans}, C.~J. 2009, \mnras, 400, 1479

\bibitem[{{Siess} {et~al.}(2000){Siess}, {Dufour}, \& {Forestini}}]{Siess}
{Siess}, L., {Dufour}, E., \& {Forestini}, M. 2000, \aap, 358, 593

\bibitem[{{Smith}(2006)}]{Smith2006}
{Smith}, N. 2006, \mnras, 367, 763

\bibitem[{{Smith} {et~al.}(2001){Smith}, {Brickhouse}, {Liedahl}, \&
  {Raymond}}]{Smi01}
{Smith}, R.~K., {Brickhouse}, N.~S., {Liedahl}, D.~A., \& {Raymond}, J.~C.
  2001, \apjl, 556, L91

\bibitem[{{Stelzer} {et~al.}(2005){Stelzer}, {Flaccomio}, {Montmerle},
  {Micela}, {Sciortino}, {Favata}, {Preibisch}, \& {Feigelson}}]{Ste05}
{Stelzer}, B., {Flaccomio}, E., {Montmerle}, T., {et~al.} 2005, \apjs, 160, 557

\bibitem[{{Stelzer} \& {Neuh{\"a}user}(2001)}]{Ste01}
{Stelzer}, B. \& {Neuh{\"a}user}, R. 2001, \aap, 377, 538

\bibitem[{{Sugitani} {et~al.}(2007){Sugitani}, {Watanabe}, {Tamura}, {Kandori},
  {Hough}, {Nishiyama}, {Nakajima}, {Kusakabe}, {Hashimoto}, {Nagayama},
  {Nagashima}, {Kato}, \& {Fukuda}}]{Sug07}
{Sugitani}, K., {Watanabe}, M., {Tamura}, M., {et~al.} 2007, \pasj, 59, 507

\bibitem[{{Townsley} {et~al.}(2011{\natexlab{a}}){Townsley}, {Broos},
  {Corcoran}, {Feigelson}, {Gagn{\'e}}, {Montmerle}, {Oey}, {Smith}, {Garmire},
  {Getman}, {Povich}, {Remage Evans}, {Naz{\'e}}, {Parkin}, {Preibisch},
  {Wang}, {Wolk}, {Chu}, {Cohen}, {Gruendl}, {Hamaguchi}, {King}, {Mac Low},
  {McCaughrean}, {Moffat}, {Oskinova}, {Pittard}, {Stassun}, {ud-Doula},
  {Walborn}, {Waldron}, {Churchwell}, {Nichols}, {Owocki}, \& {Schulz}}]{Tow11}
{Townsley}, L.~K., {Broos}, P.~S., {Corcoran}, M.~F., {et~al.}
  2011{\natexlab{a}}, \apjs, 194, 1

\bibitem[{{Townsley} {et~al.}(2011{\natexlab{b}}){Townsley}, {Broos},
  {Corcoran}, {Feigelson}, {Gagn{\'e}}, {Montmerle}, {Oey}, {Smith}, {Garmire},
  {Getman}, {Povich}, {Remage Evans}, {Naz{\'e}}, {Parkin}, {Preibisch},
  {Wang}, {Wolk}, {Chu}, {Cohen}, {Gruendl}, {Hamaguchi}, {King}, {Mac Low},
  {McCaughrean}, {Moffat}, {Oskinova}, {Pittard}, {Stassun}, {ud-Doula},
  {Walborn}, {Waldron}, {Churchwell}, {Nichols}, {Owocki}, \&
  {Schulz}}]{Town11}
{Townsley}, L.~K., {Broos}, P.~S., {Corcoran}, M.~F., {et~al.}
  2011{\natexlab{b}}, \apjs, 194, 1

\bibitem[{{Wang} {et~al.}(2008){Wang}, {Townsley}, {Feigelson}, {Broos},
  {Getman}, {Rom{\'a}n-Z{\'u}{\~n}iga}, \& {Lada}}]{Wang08}
{Wang}, J., {Townsley}, L.~K., {Feigelson}, E.~D., {et~al.} 2008, \apj, 675,
  464

\bibitem[{{Wang} {et~al.}(2007){Wang}, {Townsley}, {Feigelson}, {Getman},
  {Broos}, {Garmire}, \& {Tsujimoto}}]{Wang07}
{Wang}, J., {Townsley}, L.~K., {Feigelson}, E.~D., {et~al.} 2007, \apjs, 168,
  100

\bibitem[{{Weisskopf} {et~al.}(2002){Weisskopf}, {Brinkman}, {Canizares},
  {Garmire}, {Murray}, \& {Van Speybroeck}}]{Weis02}
{Weisskopf}, M.~C., {Brinkman}, B., {Canizares}, C., {et~al.} 2002, \pasp, 114,
  1

\bibitem[{{Winston} {et~al.}(2010){Winston}, {Megeath}, {Wolk}, {Spitzbart},
  {Gutermuth}, {Allen}, {Hernandez}, {Covey}, {Muzerolle}, {Hora}, {Myers}, \&
  {Fazio}}]{Wins10}
{Winston}, E., {Megeath}, S.~T., {Wolk}, S.~J., {et~al.} 2010, \aj, 140, 266

\bibitem[{{Wright} {et~al.}(2010){Wright}, {Drake}, {Drew}, \&
  {Vink}}]{Wright10}
{Wright}, N.~J., {Drake}, J.~J., {Drew}, J.~E., \& {Vink}, J.~S. 2010, \apj,
  713, 871

\bibitem[{{Wuchterl} \& {Tscharnuter}(2003)}]{Wuc03}
{Wuchterl}, G. \& {Tscharnuter}, W.~M. 2003, \aap, 398, 1081

\bibitem[{{Zhekov} \& {Palla}(2007)}]{Zhek07}
{Zhekov}, S.~A. \& {Palla}, F. 2007, \mnras, 382, 1124

\end{thebibliography}

\newpage


\end{document}